\documentclass[aps,showpacs]{revtex4}
\usepackage{graphicx}
\newcommand{\be}{\begin{equation}}
\newcommand{\ee}{\end{equation}}
\newcommand{\ben}{\begin{eqnarray}}
\newcommand{\een}{\end{eqnarray}}

\newcommand{\wt}{\widetilde}
\begin{document}

\title{Defect structures in sine-Gordon like models}
\author{D. Bazeia, L. Losano and R. Menezes}
\affiliation{Departamento de F\'\i sica, Universidade Federal da Para\'\i ba,
Caixa Postal 5008, 58051-970 Jo\~ao Pessoa, Para\'\i ba, Brazil}
\date{\today}

\begin{abstract}
We investigate several models described by real scalar fields, searching for topological
defects. Some models are described by a single field, and support one or two topological
sectors, and others are two-field models, which support several topological sectors.
Almost all the defect structures that we find are stable and finite energy solutions
of first-order differential equations that solve the corresponding equations of motion.
In particular, for the double sine-Gordon model we show how to find small and large BPS
solutions as deformations of the BPS solution of the $\phi^4$ model. And also, for most
of the two field models we find the corresponding integrating factors, which lead to
the complete set of BPS solutions, nicely unveiling how they bifurcate among the several
topological sectors. 

\end{abstract}
\maketitle

\section{Introduction}
\label{intro}

This work deals with kinks and domain walls in sine-Gordon like models in $(1,1)$ spacetime
dimensions. They are extended classical solutions of topological profile, supposed to
play important role in several different contexts as for instance in condensed
matter \cite{e,w} and in high energy physics \cite{r,vs}.

Models described by real scalar fields in $(1,1)$
space-time dimensions are among the simplest systems that support
topological solutions. Usually, the topological solutions are classical static solutions
of the equations of motion, with topological behavior related to the asymptotic form of
the field configurations. The topological profile can be made quantitative, with the
inclusion of the topological current, $j_T^\mu=(1/2)\varepsilon^{\mu\nu}\partial_\nu\phi.$
It gives the topological charge $Q_T=(1/2)(\phi(x\to\infty)-\phi(x\to-\infty)),$  which
is not zero when the asymptotic value of the field differs in both the positive and negative
directions. To ensure that the classical solutions have finite energy,
one requires that the asymptotic behavior of the solutions is identified
with minima of the potential that defines the system under consideration,
so in general the potential has to include at least two distinct minima
in order for the system to support topological solutions.

We can investigate real scalar fields in $(3,1)$ space-time dimensions, and
now the topological solutions are named domain walls. These domain walls are
bidimensional structures that carry surface tension, which is identified with
the energy of the classical solutions that spring in $(1,1)$ space-time
dimensions. The domain wall structures are supposed to play a role in
applications to several different contexts, ranging from the low energy scale
of condensed matter \cite{e,w,ku84,sle98} up to the high energy scale
required in the physics of elementary particles, fields and cosmology
\cite{r,ktu90,vs}.

There are at least three classes of models that support
kinks or domain walls, and we further explore such models in the next
Sec.~{\ref{gen}}. In the first class of models one deals with a single
real scalar field, and the topological solutions are structureless.
Examples of this are the sine-Gordon and $\phi^4$ models
\cite{r}. In the second class of models one also deals with a single
real scalar field, but now the systems comprise at least two distinct
domain walls. An example of this is the double sine-Gordon model,
which has been investigated for instance in Refs.~{\cite{ma,dst80,cgm83,cps86,dmu,mrs}}.
In the third class of models we deal with systems defined by two real scalar
fields, where one finds domain walls
that admit internal structure \cite{mke,mor,brs,97,98,mor98,bbb99},
and junctions of domain walls, which appear in models of two fields when
the potential contains non-collinear minima, as recently investigated for
instance in Refs.{\cite{99a,99b,99c,00a,00b,00c,agm,h,00d,bv01,bb01,n,nns01}}.
We study new possibilities in Sec.~{\ref{cla}}, and there we investigate
periodic systems, which present several topological sectors, which may bifurcate
into richer structures. For most of them, we find the integrating factors,
which lead us to all the BPS solutions the models engender.

There are other motivations to investigate
domain walls in models of field theory, one of them being related to
the fact that the low energy world volume dynamics of branes in string and
M theory may be described by standard models in field
theory \cite{s1,s2,s3}. Besides, one knows that field theory models
of scalar fields may also be used to investigate properties of quasi-linear
polymeric chains, as for instance in the applications of
Refs.~{\cite{96,99,00,01}}, to describe solitary waves in ferroelectric
crystals, the presence of twistons in polyethylene,
and solitons in Langmuir films.

Domain walls have been observed in several
different scenarios in condensed matter, for instance in ferroelectric
crystals \cite{sle98}, in one-dimensional nonlinear lattices \cite{prl},
and more recently in higher spatial dimensions -- see \cite{nat} and references
therein. The potentials that appear in the models of field theory that we
investigate in the present work are also of interest to map systems described
by the Ginzburg-Landau equation, since they may be used to explore the
presence of fronts and interfaces that directly contribute to pattern
formation in reaction-diffusion and in other spatially extended, periodically
forced systems \cite{ku84,w,cou90,ce90,ehm98,be98}.

For completeness, in Sec.~{\ref{nbps}} we investigate other topological sectors of the periodic
two field models, where the domain wall solutions are of the non-BPS type. And finally, we end
the paper in Sec.~{\ref{con}}, where we include our comments and conclusions.

\section{Single field models}
\label{gen}

In this work we are interested in field theory models that describe real
scalar fields and support topological solutions of the
Bogomol'nyi-Prasad-Sommerfield (BPS) type \cite{b,ps}.
In the case of a single real scalar field $\phi$, we consider
the Lagrange density
\be
\label{1f}
{\cal L}=\frac12\partial_{\mu}\phi\partial^{\mu}\phi-V(\phi)
\ee
Here $V(\phi)$ is the potential, which identifies the particular model
under consideration. We write the potential in the form
\be \nonumber
V(\phi)=\frac12W^2_{\phi}
\ee
where $W=W(\phi)$ is a smooth function of the field
$\phi$, and $W_\phi=dW/d\phi$. In a supersymmetric theory $W$ is the
superpotential, and this is the way we name $W$ in this work.

The equation of motion for $\phi=\phi(x,t)$ has the general form
\be
\label{em2}
\frac{\partial^2\phi}{\partial t^2}-
\frac{\partial^2\phi}{\partial x^2}+\frac{dV}{d\phi}=0
\ee
and for static solutions we get
\be
\label{ems}
\frac{d^2\phi}{dx^2}=W_{\phi}W_{\phi\phi}
\ee
It was recently shown in Refs.~{\cite{bms01a,bms01b}} that this equation
of motion is equivalent to the first order equations
\be
\label{emf}
\frac{d\phi}{dx}=\pm W_{\phi}
\ee
if one is searching for solutions that obey the boundary conditions
$\lim_{x\to-\infty}\phi(x)={\bar\phi}_i$ and
$\lim_{x\to-\infty}(d\phi/dx)=0$, where ${\bar\phi}_i$ is one among the
several vacua $\{{\bar\phi}_1,{\bar\phi}_2,...\}$ of the system.
In this case the topological solutions are BPS (or anti-BPS) states.
Their energies get minimized to the value $t^{ij}=|\Delta W_{ij}|$,
where $\Delta W_{ij}=W_i-W_j$, with $W_i$ standing for $W({\bar\phi}_i)$.
The BPS solutions are defined by two vacuum states belonging
to the set of minima that identify the several topological sectors of
the model.

In a recent work \cite{blm02} one has introduced a deformation prescription,
in which one defines a deformation function $f=f(\phi),$ which allows
introducing new models presenting defect solutions. The deformation prescription
goes as follows: if we change the potential $V(\phi)$ to ${\wt V}(\wt\phi)$,
given by
\be \nonumber
{\wt V}(\wt\phi)=V(\phi\to f)/(d\, f/d{\wt\phi})^2
\ee
where $f=f(\wt\phi)$ is a well-defined, invertible function, then the defect
solution of the new model can be obtained as
\be \nonumber
{\wt\phi}(x)=f^{-1}(\phi(x))
\ee
where $\phi(x)$ represents the defect solution of the former model. More recently,
in another work \cite{ablm} one has enlarged the scope of the procedure,
using other deformations.

\subsection{Single field. One kink solutions}
\label{ofp1}

We now turn attention to kinks and domain walls. Perhaps the most
known example of this is given by the $\phi^4$
model, defined by the potential $V(\phi)=\frac12\lambda^2\,(v^{2}-\phi^2)^2$,
where we are using natural units.
In this model the domain wall can be represented by the solution
$\phi_s(x)=\pm v\tanh(\lambda v x)$. The above potential can be written
with the superpotential $W(\phi)=\lambda(v^{2}\phi-\phi^3/3)$, and the domain
wall is of the BPS type. The wall tension corresponding to the BPS kink
is $t_s=4/3|\lambda v^3|$.

Another example of system described by one field can be generated from the above
$\phi^4$ model, using the deformation method \cite{blm02,ablm} with the deformation
function $f(\phi)=v\sin(v\phi)$. In this case we obtain the sine-Gordon model
defined by the potential
\be
\label{v1}
V=\frac12\lambda^{2}\cos^{2}(v\phi)
\ee
with the kink and antikink solutions
\be
\label{fi2}
\phi(x)=\pm\frac{1}{v} \arcsin\bigl[\tanh(\lambda v x)\bigr]+k\pi
\ee
This potential can be written with the
superpotential $W(\phi)=(\lambda/v)\,\sin(v\phi),$ and the kink is of the BPS type.
There is an infinite set of minima given by
$\phi= k\,\pi/v$, were k is positive or negative integer.
The wall tension corresponding to the BPS wall is
$t_s=2|\lambda/v|$.

\subsection{Single field. Two kink solutions}
\label{ofp2}

We can build another class of models, in which the domain walls
engender other features. We refer to models described by a single field,
but the systems may now support two or more distinct topological solutions.
An interesting example of this is the double sine-Gordon model defined by the potential,
\be
\label{dsgg}
V=\frac{\lambda^{2}}{2\alpha^{2}v^{2}}
\bigl[\cos^{2}(v\phi)-\alpha^{2}\,\sin^{2}(v\phi)\bigr]^{2}
\ee
which can be generated from the $\phi^4$ model above, employing the deformation
function $f_1(\phi)=\alpha\tan(v\phi)$ or
$f_2(\phi)=\frac{1}{\alpha}\tan(v\phi-\frac{\pi}{2})$. It is interesting
to see that if one uses the deformation prescription with the $f_1$-deformation,
we get the solutions 
\be
\label{phi1}
\phi_1(x)=\pm\frac{1}{v}\arctan\left(\frac{1}{\alpha}\tanh(\lambda x)\right)+
\frac{n\pi}{v}
\ee
However, if one uses the $f_2$-deformation there are new solutions, given by
\be
\label{phi2}
\phi_2(x)=\mp\frac{1}{v}\arctan\left(\alpha\tanh(\lambda x)\right)+
\frac{(2n+1)\pi}{2v}
\ee
These solutions represent large kinks and small kinks,
which appear in the double sine-Gordon model. The novelty here is that we are finding
the large and small kink solutions, using the deformation procedure of
Refs.~\cite{blm02,ablm}. In Fig.~1, we depict the behavior of the potential in
Eq.(\ref{dsgg}), in terms of the parameter $\alpha$. There we see the double sine-Gordon
models, and we also show the topological sectors corresponding to the solutions
$\phi_1(x)$ and $\phi_2(x).$

\begin{figure}[ht]
\includegraphics[{height=6cm}]{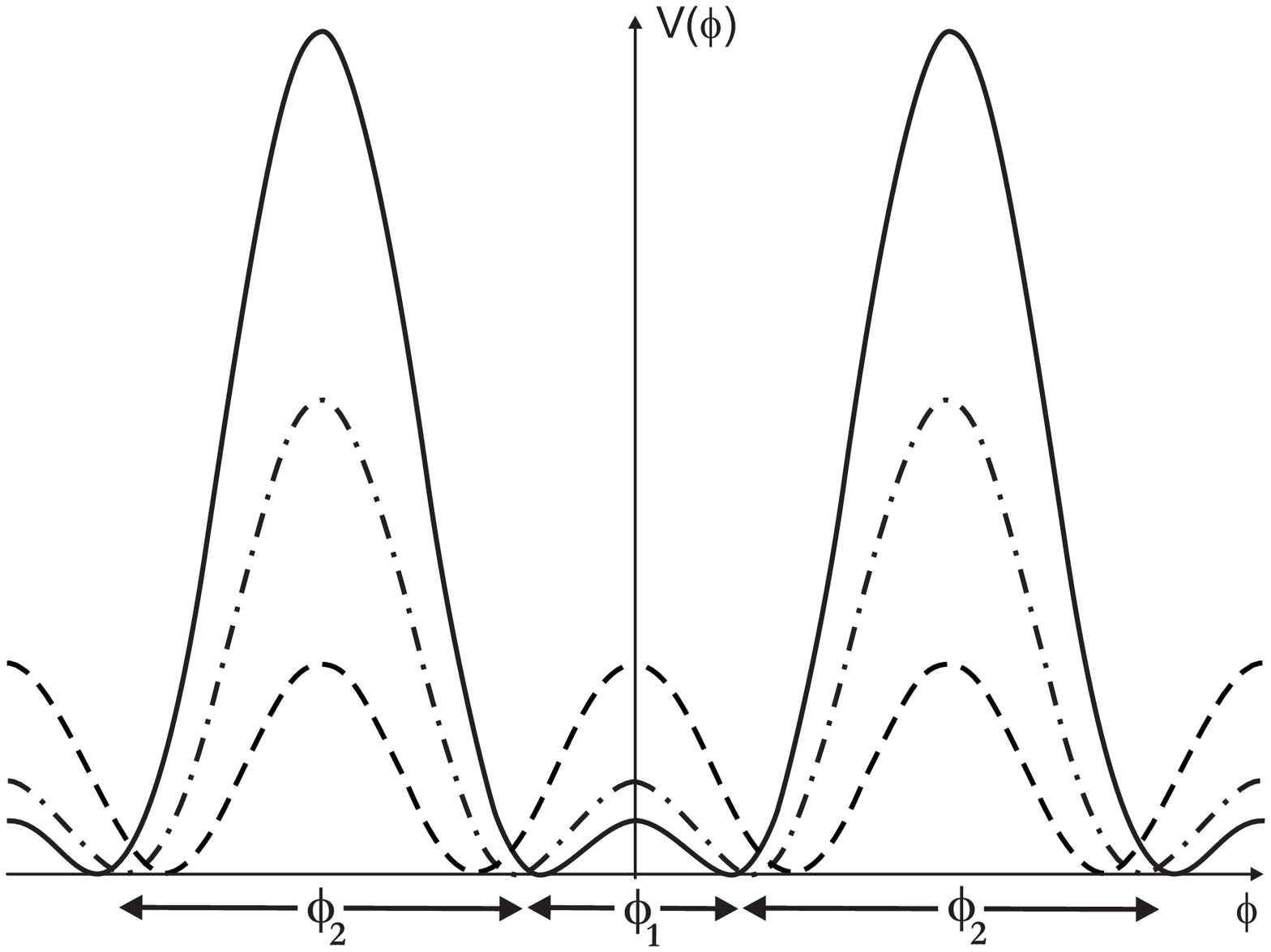}
\includegraphics[{height=6cm}]{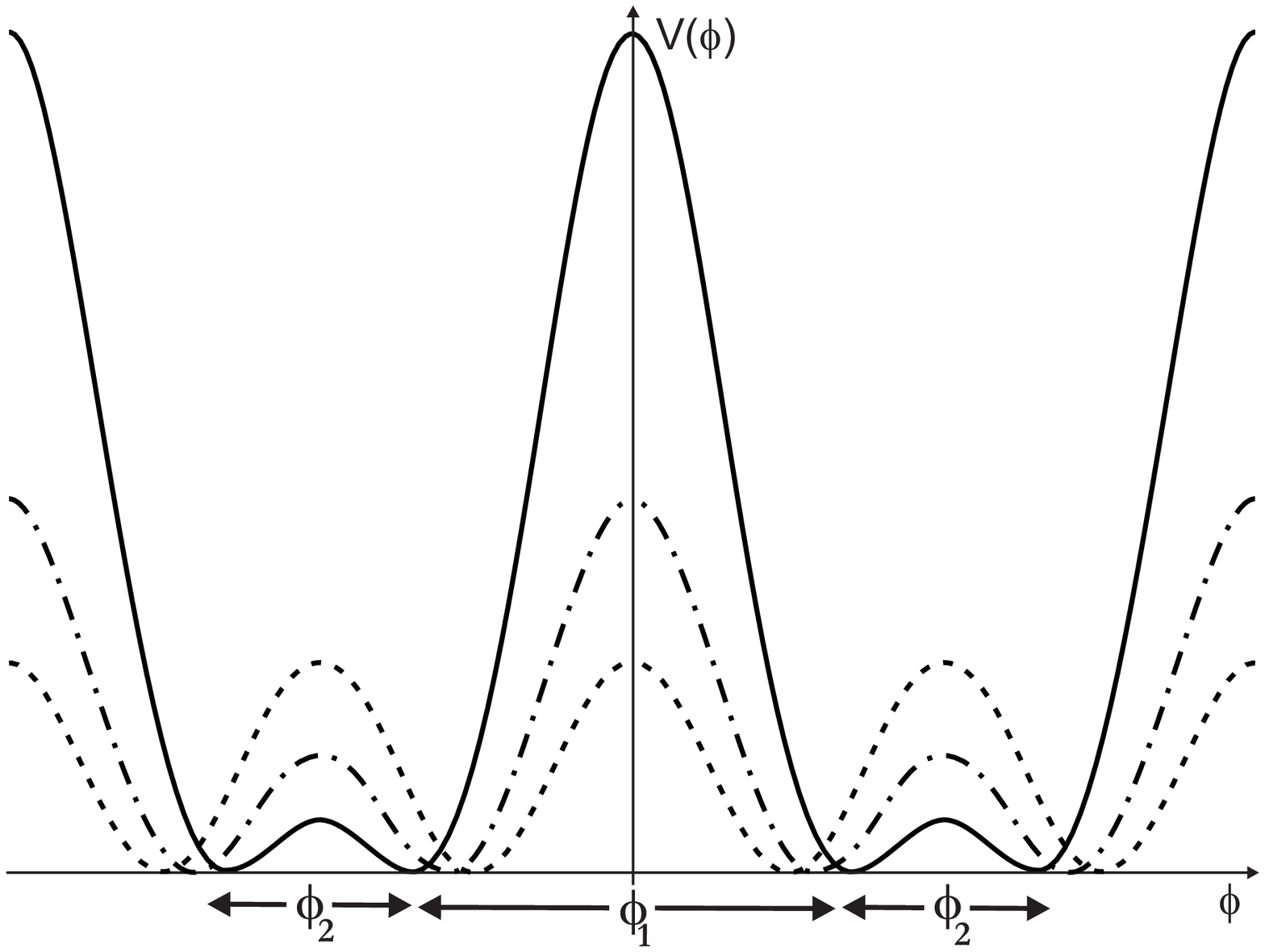}
\vspace{0.3cm}
\caption{The double sine-Gordon model. The left figure shows
the case $\lambda=v=1,$ for $\alpha=1$ (dashed line), $\alpha=1.5$
(dash-dotted line), and $\alpha=2$ (solid line). The right figure shows
the case $\lambda=v=1$ for $\alpha=1$ (dotted line), $\alpha=0.75$
(dash-dotted line), and $\alpha=0.50$ (solid line).}
\end{figure}

The potential (\ref{dsgg}) does not give double sine-Gordon behavior as $\lambda$
and $v$ vary, for $\alpha=1.$ Thus, since we shall concentrate the investigation on the double
sine-Gordon model, which profile depends only on $\alpha,$ we prefer to change the
potential to the form
\be
\label{dsg}
V_r(\phi)=\frac1{r+1}\bigl[4\,r\,\cos(\phi)+\cos(2\phi)\bigr]
\ee
where $r$ is a parameter, real and positive. This potential is periodic,
with period $2\pi$, and for simplicity in the following we consider the
interval $-2\pi<\phi<2\pi$. The value $r=1$ distinguishes two regions,
the region $r\in(0,1)$ where the potential contains four minima,
and the region $r\geq1$, where the potential contains two
minima. For $r\in(0,1)$ the system supports two distinct
wall configurations, the large wall and the small wall, which distinguish
the two different barrier the model comprises in this case. The limits
$r\to 0$ and $r\to\infty$ lead us back to the sine-Gordon model, with period
$\pi$ and $2\pi$, respectively. The double sine-Gordon model has been considered
in several distinct applications, as for instance in
Ref.~{\cite{ma,dst80,cgm83}}, where one investigates magnetic solitons
in superfluid $^3$He, kink propagation in a model for poling in polyvinylidine
fluoride, and properties related to the two different kinks that appear in such
polymeric chain. More recently, it has also been used to model magnetic solitons
in uniaxial antiferromagnetic systems \cite{brwm}. 

To expose new features of the double sine-Gordon model we rewrite
Eq.~(\ref{dsg}) in the form
\be
\label{dsgw}
V_r(\phi)=\frac2{1+r}[\cos(\phi)+r\,]^2
\ee
where we have omitted an unimportant r-dependent constant.
This potential is a particular case of the potential (\ref{dsgg}),
with $\alpha=\sqrt{(1-r)/(1+r)},\; v=1/2$, and $\lambda=\sqrt{1-r}$.
Thus, it is a deformation of the $\phi^4$ model, and can be described
by the superpotential
\be \nonumber
W(\phi)=\frac{2}{\sqrt{1+r}}[\sin(\phi)+r\,\phi]
\ee

\begin{figure}[ht]
\includegraphics[{height=6cm}]{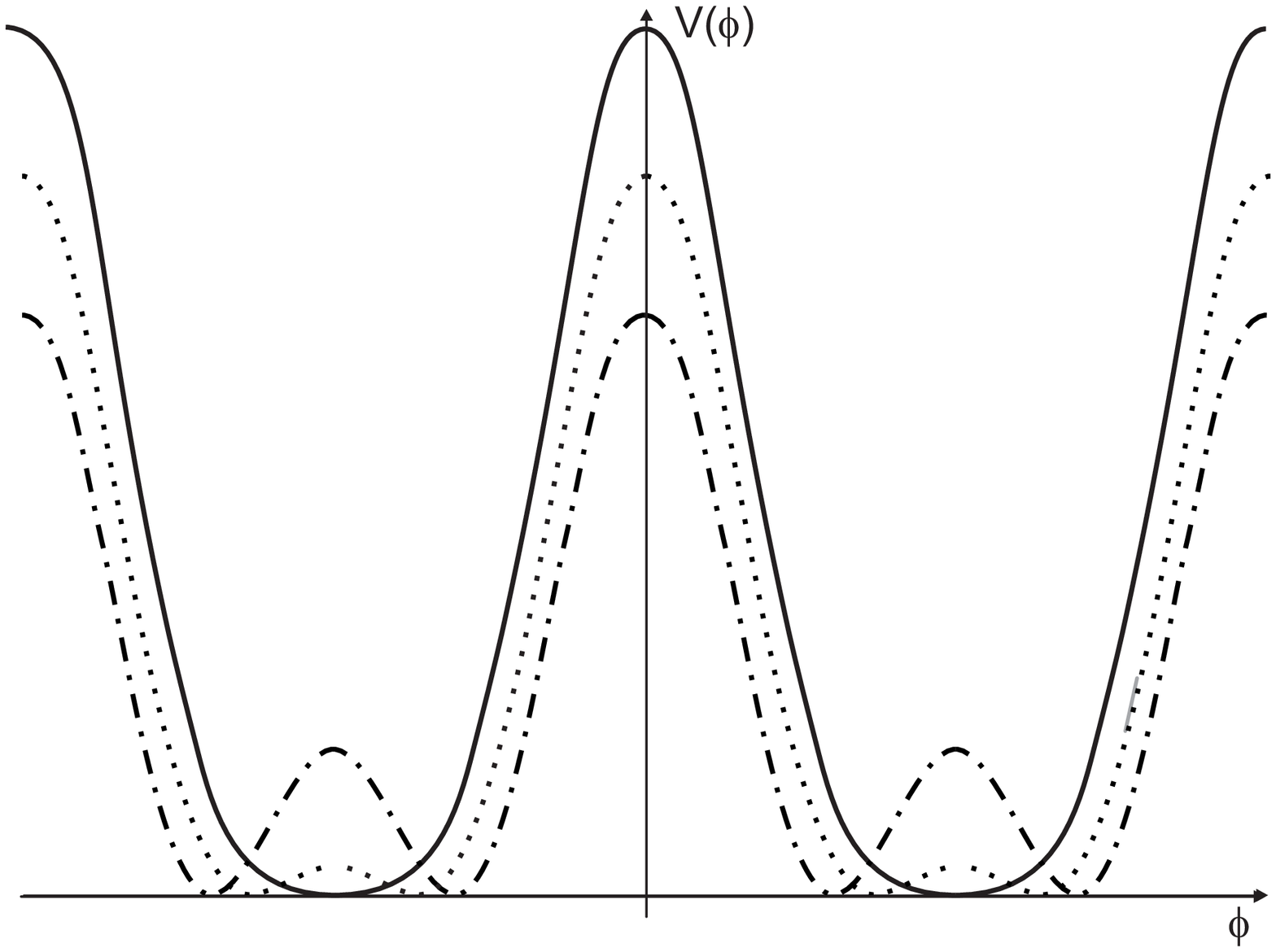}
\vspace{0.3cm}
\caption{The double sine-Gordon potential, depicted for $r=1/3$ (dash-dotted line),
$r=2/3$ (dotted line), and $r=1$ (solid line), to illustrate how the behavior
of the model changes with $r$. Our notation is such that both $V(\phi)$ and $\phi$
are dimensionless, as we explain in the work.}
\end{figure}

For $r$ in the interval $r\in(0,1)$ the minima of the potential
are the singular points of the superpotential, $dW/d\phi=0$. They are
periodic, and for $-2\pi<\phi<2\pi$ there are four minima, at the points
${\bar\phi}=\pm\pi\pm\,\alpha(r)$,
where $\alpha(r)=\arccos(r)$. For $r\geq1$ the minima
are at ${\bar\phi}=\pm\pi$, in the interval $-2\pi<\phi<2\pi$. A closer
inspection shows that for $0<r<1$ the local maxima at $\pm\pi$
and the minima $\pm\pi\pm\alpha(r)$ degenerate to the minima
$\pm\pi$ for $r=1$, and remain there for $r>1$.
Thus, the parameter $r$ induces a transition in the
behavior of the double sine-Gordon model.

The value $r=1$
is the critical value, since it is the point where the system changes behavior:
for $r\in(0,1)$ this model supports minima that do {\it not} appear for $r\geq1$.
We illustrate the double sine-Gordon model in Fig.~[2], where we depict the
potential (\ref{dsgw}) for $r=1/3,2/3$ and for $r=1$. Thus, it is very interesting
to see that this model maps the presence of magnetic solitons in uniaxial systems
\cite{brwm}: in fact, the model is able to model both the antiferromagnetic, canted and
weak ferromagnetic phases. In particular, in the canted phase there are two different
domain walls connecting minima between two different barriers, as we show below.  

To get a better view of the the double sine-Gordon model, we examine the order
parameter ${\bar\phi}(r)$, which is given by $ \pm\pi\pm\,\alpha(r)$ for
$0<r\leq1$, so it goes continuously to $\pm\pi$ for $r\geq1$. Also, the (squared)
mass of the field can be obtained via the relation
\be
V_r''({\bar\phi})= W^2_{{\bar\phi}{\bar\phi}}+
W_{{\bar\phi}}W_{{\bar\phi}{\bar\phi}{\bar\phi}}
\ee
where ${\bar\phi}$ is the corresponding minimum of the potential. For
$0<r\leq1$ we get $m^2(r)=4-4r$, and for $r\geq1$ we have
$m^2(r)=4(r-1)/(r+1)$. We see that $m(r)$ vanishes in the limit $r\to1$.
These results indicate that $r$ mimics a second order phase transition,
a transition where the system goes from the case of two distinct phases
to another one, engendering a single phase, as one illustrates in Fig.~3.

\begin{figure}[ht]
\includegraphics[{height=6cm}]{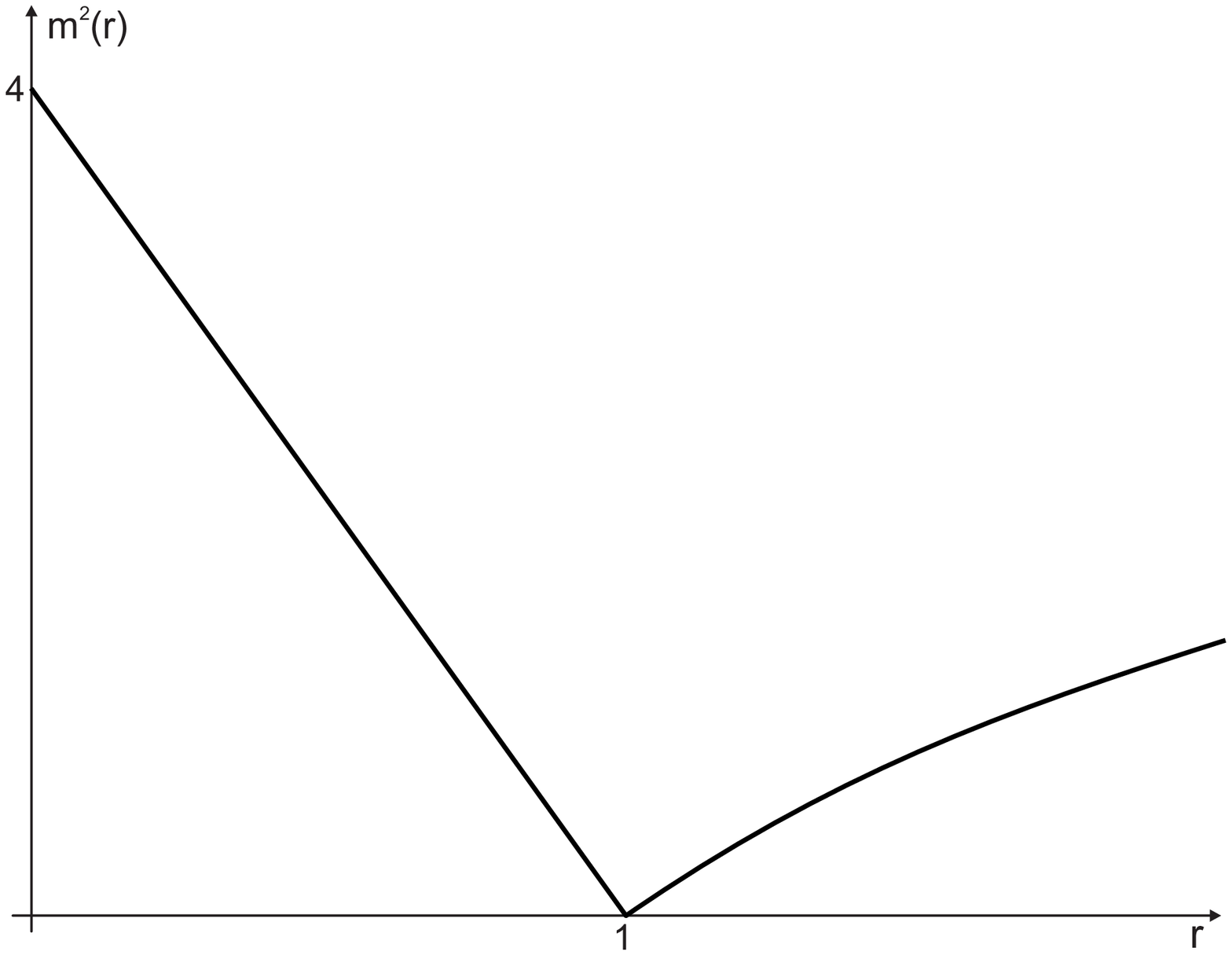}
\includegraphics[{height=6cm}]{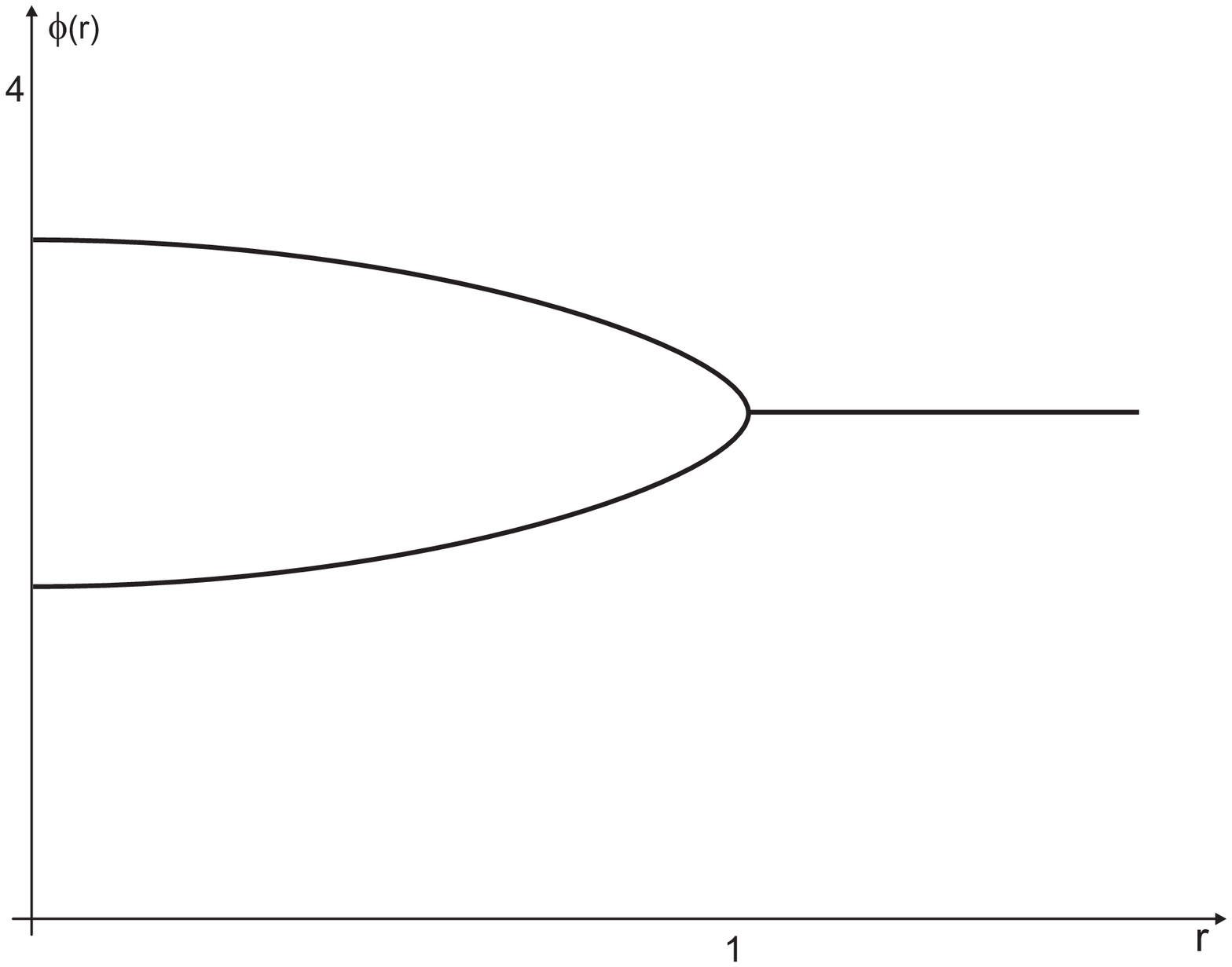}
\vspace{0.3cm}
\caption{Plots of $m^2(r)$ (left) and ${\bar{\phi}}(r)$ (right) for the
double sine-Gordon potential, which illustrate how the behavior of the
model changes with $r$. The quantities ${\bar{\phi}}$, $m(r)$,
and $r$ that appear in these figures are all dimensionless.}
\end{figure}

Let us first consider the case $0<r\leq1$. The energies of the BPS solutions are
given as follows. For solutions connecting the minima $-\pi+\alpha(r)$ and
$\pi-\alpha(r)$ the defect is large since it joins minima separated by a higher
and wider barrier. We have
\be \nonumber
t^l_{dsG}=4\sqrt{1-r}+4r\frac{\pi-\alpha(r)}{\sqrt{1+r}}
\ee
In the case of the minima $\pi-\alpha(r)$ and $\pi+\alpha(r)$ the defect
is small and we get
\be \nonumber
t^s_{dsG}=4\sqrt{1-r}-4r\frac{\alpha(r)}{\sqrt{1+r}}
\ee

We notice that $t^l_{dsG}=t^s_{dsG}+4\pi r/\sqrt{1+r}$, and the limit
$r\to1$ sends $t^l_{dsG}\to 2\sqrt{2}\pi$ and $t^s_{dsG}\to0$, as expected.
For the BPS states, from Eqs.~(\ref{phi1}) and (\ref{phi2}) we can write the
solutions explicitly. For instance, for solutions that connect the minima
$-\pi+\alpha(r)$ and $\pi-\alpha(r)$ we get large kink solutions,
which are of the form
\be 
\phi_{l}(x)=\pm 2\arctan\Biggl[\sqrt{\frac{1+r}{1-r}}\,
\tanh\left(\sqrt{1-r}\;x\right)\Biggr]
\ee
For solutions that connect the minima $\pi\pm\alpha(r)$ and the minima
$-\pi\pm\alpha(r)$ we get small kink solutions. They are given by
\be
\phi_{s}(x)=\pm\pi-2\arctan\Biggl[\sqrt{\frac{1-r}{1+r}}\,
\tanh\left(\sqrt{1-r}\,x\right)\Biggr]
\ee
The case $r>1$ is different. The minima are now at $\pm\pi$, and the model
is similar to the standard sine-Gordon model. In this case we modify
the potential of Eq.~(\ref{dsgw}) to
\be \nonumber
V(\phi)=\frac{2}{r+1}(\cos^2(\phi)+2r \cos(\phi)+2r-1)
\ee
to make $V(\pm\pi)=0$. Thus, we can write this potential in terms
of another superpotential, such that
\be \nonumber
{\widetilde W}_{\phi}=\frac{2}{\sqrt{r+1}}\sqrt{\cos^2(\phi)+
2r \cos(\phi)+2r-1}
\ee
The presence of the square root complicates the calculation, and we have been
unable to find explicit analytical solutions in this case.

The potential in Eq.~(\ref{dsgw}) in the limit $r\to0$ goes to $V_0(\phi)=1+\cos(2\phi)$
which leads us back to the sine-Gordon model. Thus, we can suppose
$r$ small and use $V_r(\phi)$ to explore the double sine-Gordon
model as a model controlled by a small parameter, in the vicinity
of the sine-Gordon model. This feature is of direct interest to investigations
that follow the lines of Ref.~{\cite{abl01}}, which explores the vicinity
of the BPS bound, and also in the case concerning the presence of internal
modes of solitary waves, which seems to appear when one slightly modifies
integrable models -- see for instance Ref.~{\cite{k98}}.

\section{Two field models}
\label{cla}

We now turn attention to another class of models, which is described
by two real scalar fields. In this case the domain walls may engender internal
structure; see, e.g., Refs.~{\cite{mke,mor,brs}}. In the present work we are
interested in models of the sine-Gordon type, but now described by two real scalar
fields.

In the case of two real scalar fields $\phi$ and $\chi,$ the potential
is written in terms of the superpotential, in a way such that
\be
V(\phi,\chi)=\frac12\,W^{2}_{\phi}+\frac12\,W^{2}_{\chi}
\ee
The equations of motion for static fields are
\be \nonumber
\frac{d^{2}\phi}{dx^{2}}=W_{\phi}W_{\phi\phi}+W_{\chi}W_{\chi\phi}
\ee
\be \nonumber
\frac{d^{2}\chi}{dx^{2}}=W_{\phi}W_{\phi\chi}+W_{\chi}W_{\chi\chi}
\ee
which are solved by the first order equations
\be
\frac{d\phi}{dx}=\pm W_{\phi}\,\;\;\;\;\;\;\;\;
\frac{d\chi}{dx}=\pm W_{\chi}
\ee
Solutions to these first order equations constitute the BPS states of the model.
They solve the equations of motion, and have energy minimized to
$t^{ij}=|\Delta W^{ij}|$ as in the case of a single field; here, however,
$\Delta W^{ij}=W(\phi_i,\chi_i)-W(\phi_j,\chi_j)$, since now we need
the pair $(\phi_i,\chi_i)$ to represent each one of
the vacuum states in the system of two fields. An important issue is that
in the plane $(\phi,\chi)$ one may have minima that are noncolinear, and this opens
the possibility for junctions of defects. Moreover, in the case of two real
scalar fields we can find a family of first-order equations that are equivalent
to the pair of second-order equations of motion, but this requires that
$W_{\phi\phi}+W_{\chi\chi}=0$, in the case of harmonic superpotentials
\cite{bms01a,bms01b}.

Models described by two real scalar fields are more intricate, and require
more involved investigations. For instance, to solve the two first-order equations
one can search for the integrating factor \cite{spain}. However, the
integrating factor is in general hard to find, and so we can use the trial
orbit method, as suggested in \cite{bflr} in detail.

The models that we investigate are of interest both to ferromagnetic
\cite{brwm} and ferroelectric \cite{pmo} systems. They are defined by the
following general potential
\ben
\label{v3}
V(\phi,\chi)=\frac12\eta^{2}\,\bigl\{\bigl[r\,+\,
\cos(\phi)\cos(p\,\chi)\bigr]^{2}
+\,p^{2}\sin^{2}(\phi)\sin^{2}(p\,\chi)\bigr\}
\een
where the parameter $p$ is real and positive, with $p\not=1$ [the case $p=1$
leads to single field models with a $\pi/2$ rotation in the $(\phi,\chi)$ plane].
We will consider three distinct models, with $r=0,$ $0<r<1, $ and $r\geq 1$.
The case $r=0$ and $r\geq1$ describe two coupled sine-Gordon models,
and the case $0<r<1$ describe two coupled double sine-Gordon models.
These models are new, and some of their features will be examined below.

In general, all the three models can be described by the following superpotential
\be
\label{w3}
W(\phi,\chi)=\eta\bigr[r\phi\,+\,\sin(\phi)\cos(p\,\chi)\bigl]
\ee 
and this will help simplifying the investigation, since the BPS solutions satisfy
first order differential equations, which are simpler to solve compared to the
equations of motion.

In this section we search for BPS solutions in the models introduced
in this paper. We investigate the first order equations
corresponding to each one of the three models separately. For model 1 and model 2 we obtain
the integrating factors, which allow solving the first order equations completely,
finding all the family of orbits which connect the minima of the potential with energy
minimized to BPS states.  

\subsection{Model 1}
\label{m1}

The first model is given by the potential (\ref{v3}) with $r=0.$
In terms of the superpotential (\ref{w3}) the first order equations
are
\ben
\label{dif1}
\frac{d\phi}{dx}&=&\eta\,\cos(\phi)\cos(p\,\chi)
\\
\label{dif2}
\frac{d\chi}{dx}&=&-\eta\,p\,\sin(\phi)\sin(p\,\chi)
\een
There are minima at the points 
$v_{n,m}=\bigl((2n+1)\,\pi/2\,,\,m\,\pi/p\bigr)$ and
$w_{n,m}=\bigl(n\,\pi\,,\,(2m+1)\,\pi/2p\bigr)$.
In Fig.~4 we display these minima in the plane $(\phi,p\,\chi)$. 
\begin{figure}[ht]
\includegraphics[{height=6cm}]{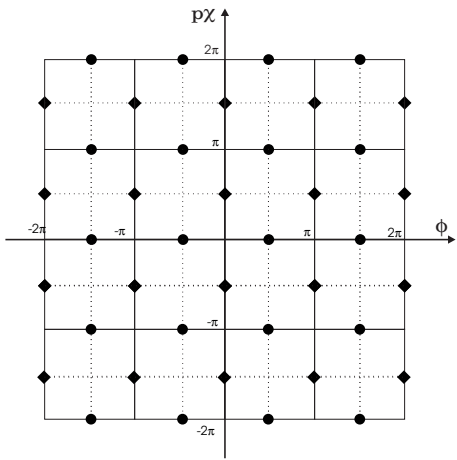}
\vspace{0.3cm}
\caption{The minima of model 1 potential. The minima $v$ are
represented by balls, and $w$ by losangles.}
\end{figure}

To obtain the most general solution to the first-order Eqs.~(\ref{dif1})
and (\ref{dif2}), we integrate the ordinary differential equation
\be \nonumber
\frac{d\phi}{d\chi}=-\frac{1}{p}\cot(\phi)\cot(p\,\chi)
\ee
which admits the integrating factor $\sin^{-\sigma}(p\,\chi)$,
with $\sigma=1+1/p^2$, and this determines all the orbits connecting the minima
via the family of curves
\be
\label{orb1}
\cos(\phi)=C\frac{1}{p}\sin^{\sigma-1}(p\,\chi)
\ee
where $C$ is a real integration constant. In fig.~5, the behavior of a particular
orbit with $C$, in the family (\ref{orb1}) is displayed, where the critical value
is $C=p$.

\begin{figure}[ht]
\includegraphics[{height=6cm}]{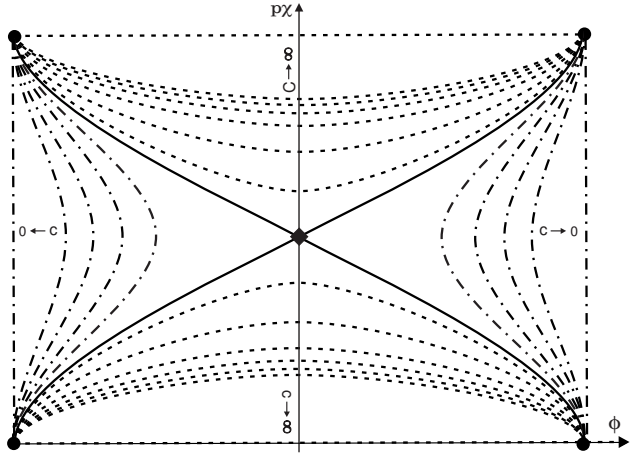}
\vspace{0.3cm}
\caption{Orbits from model 1 given for: $0\leq C<p$ (dash-dotted line),
$C=p$ (solid line), and  $p<C\leq \infty$ (dashed line).}
\end{figure}

There are three type of BPS sectors connecting pairs of adjacent minima,
namely: ${v}{v}_{1}$ for $0\leq C<p$, ${v}{w}$ for $C=p$,
and ${v}{v}_{2}$ for $p<C\leq\infty$.
The tensions of the BPS states are 
$t_{vv_{1}}=t_{vv_{2}}= 2|\eta|$, and $t_{vw}=|\eta|$.
Using the curve (\ref{orb1}), the BPS equations (\ref{dif1}) and (\ref{dif2})
decouple and can be rewritten as, for $C\in(0,\infty)$
\ben
\label{dif3}
\frac{d\phi}{dx}&=&\eta\,\cos(\phi)
\bigl[1-D^{2\nu}\,\cos^{2\nu}(\phi)\bigr]^{\frac12}
\\
\label{dif4}
\frac{d\chi}{dx}&=&-\eta\,p\,\sin(p\,\chi)
\bigl[1-E^{2/\nu}\,\sin^{2/\nu}(p\,\chi)\bigr]^{\frac12}
\een
where $D=1/C p$, $E=C^{2}/p^{2}$, and $\nu=p^{2}$.
In general, we cannot obtain the analytical solutions of the equations
(\ref{dif3}) and (\ref{dif4}). For this reason, we present some numerical
solutions for each kind of orbit, in the following.

Firstly, we investigate the existence of kinks in the ${v}{v}_{1}$ sector. For simplicity,
we consider $C=0$, which leads to orbits in the form of straight line segments connecting
two vertical adjacent ${v}$ minima, with $\phi=(2k+1)\pi/2$. This reduces the potential
(\ref{v3}) to 
\be
\label{vc1}
V=\frac12\eta^2\,p^2\,\sin^{2}(p\,\chi)
\ee
The BPS equations (\ref{dif1}) and (\ref{dif2})
have the kink and antikink solutions:
\ben
\phi_{vv_{1}}(x)&=&(2k+1)\frac{\pi}{2} \nonumber\\
\chi_{vv_{1}}(x)&=&\pm\frac{1}{p} \arccos(\pm\tanh(\eta\,p^2 x))+k\frac{\pi}{p} \nonumber
\een
In Fig.~6, we plot generic numerical solutions for $0<C<p$.

\begin{figure}[ht]
\includegraphics[{height=6cm}]{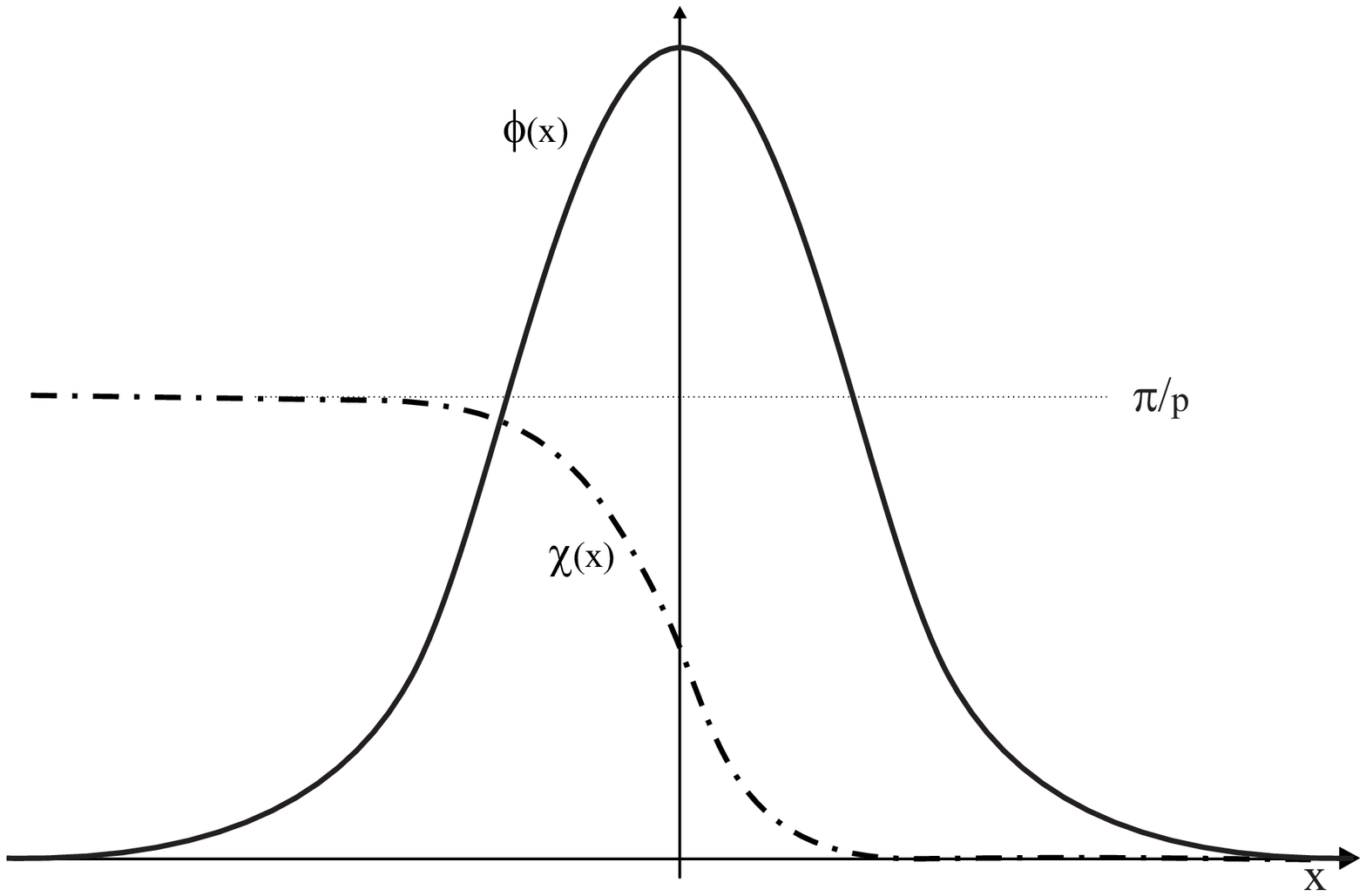}
\vspace{0.3cm}
\caption{Numerical solutions for ${v}{v}_{1}$ sector.}
\end{figure}

In the ${v}{w}$ sector, the numerical solutions for $C=1/p$ are depicted in Fig.~7.

\begin{figure}[ht]
\includegraphics[{height=6cm}]{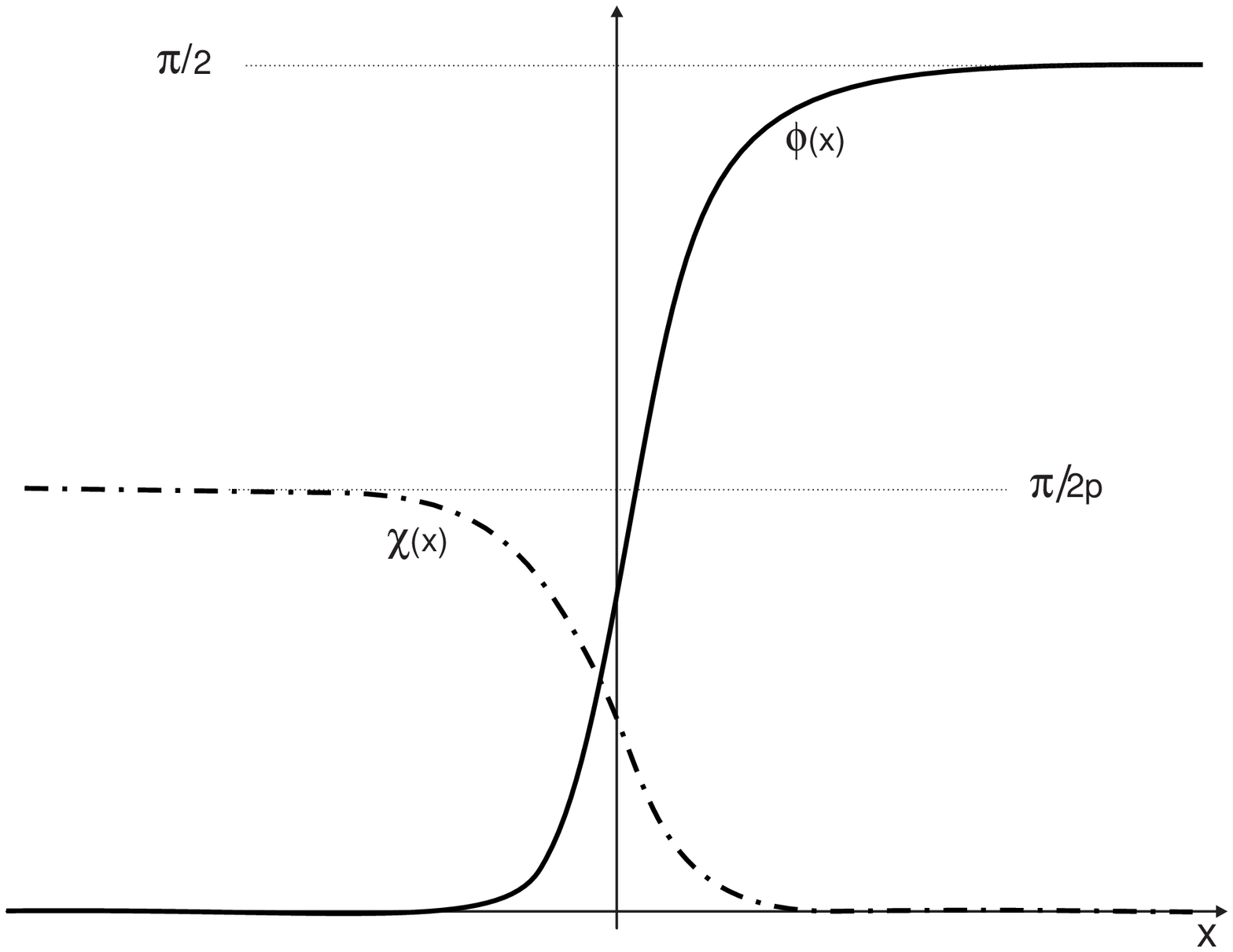}
\vspace{0.3cm}
\caption{Numerical solutions for ${v}{w}$ sector.}
\end{figure}

Finally, we examine the ${v}{v}_{2}$ sector. Two horizontal adjacent ${v}$ minima are
connected by the straight line orbit $p\chi=k\,\pi$, for $C=\infty$. In this case,
the potential (\ref{v3}) reduces to the form of Eq.~(\ref{v1}). Thus, the kink and
antikink solutions of the BPS equations are
\ben
\chi_{vv_{2}}(x)&=&k\,\frac{\pi}{p} \\
\phi_{vv_{2}}(x)&=&\pm \arcsin(\tanh(\eta\, x))+k\pi
\een

In Fig.~8, we show numerical solutions in the more general case,
for $p<C<\infty$.

\begin{figure}[ht]
\includegraphics[{height=6cm}]{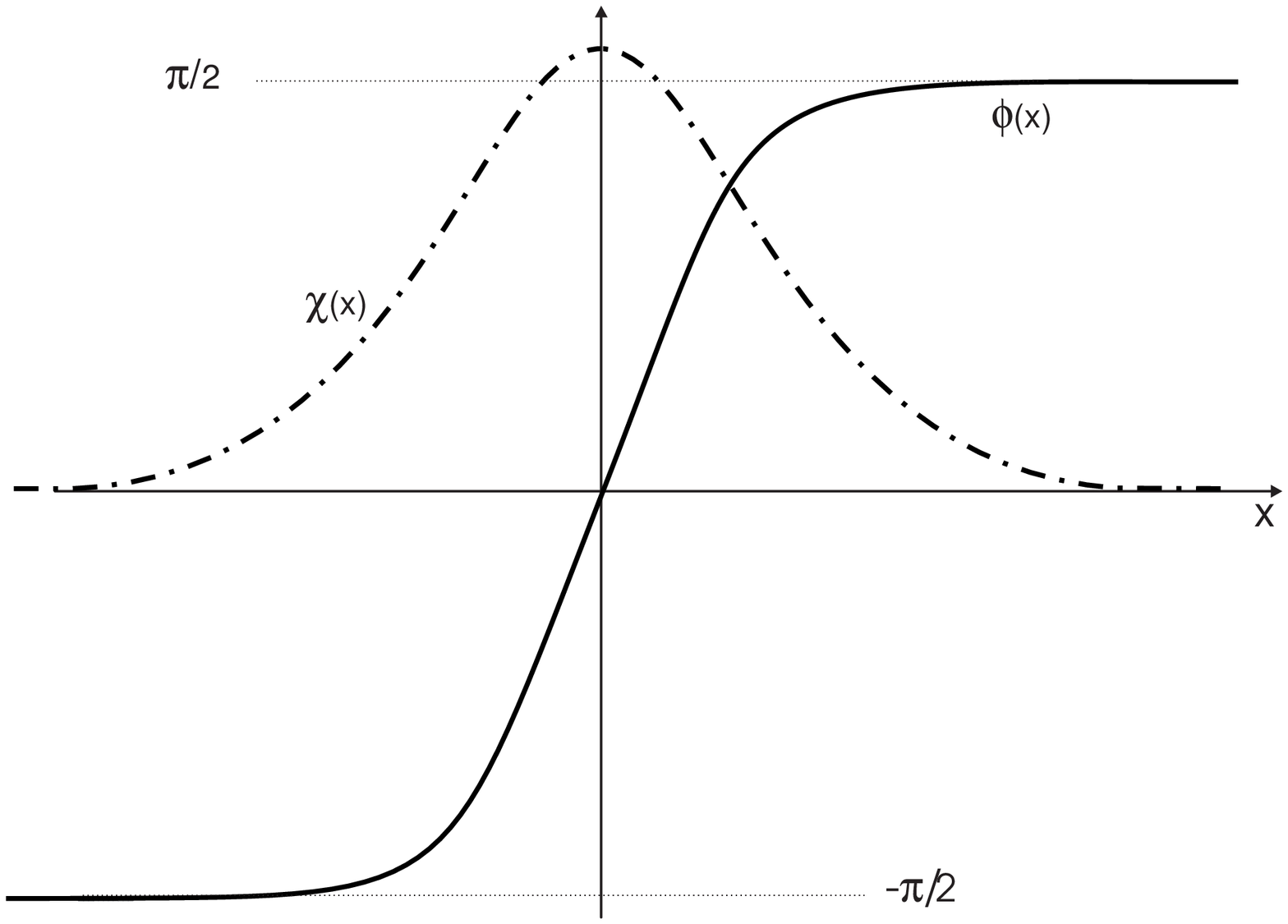}
\vspace{0.3cm}
\caption{Numerical solutions for ${v}{v}_{2}$ sector.}
\end{figure}

\subsection{Model 2}
\label{m2}

Now we explore the second model, described by the potential (\ref{v3})
with $0<r<1$. From the superpotential (\ref{w3}) the first-order equations are
\ben
\label{dm21}
\frac{d\phi}{dx}&=&\eta\,\bigl[r+\cos(\phi)\cos(p\,\chi)\bigr]
\\
\label{dm22}
\frac{d\chi}{dx}&=&-\eta\,p\,\sin(\phi)\sin(p\,\chi)
\een
The set of minima are given by 
$v_{n,m}=\bigl(\pm[(1+(-1)^{m})\pi/2-\alpha]/p+2n\pi,\,m\,\pi/p\bigr)$
and $w_{n,m}=\bigl(n\,\pi\,,\,\pm[(1+(-1)^{n})\pi/2-\alpha]/p+2m\pi/p)$,
with $\alpha=\arccos(r)$. In Fig.~9 we depict the minima in the plane
$(\phi,p\,\chi)$.
\begin{figure}[ht]
\includegraphics[{height=6cm}]{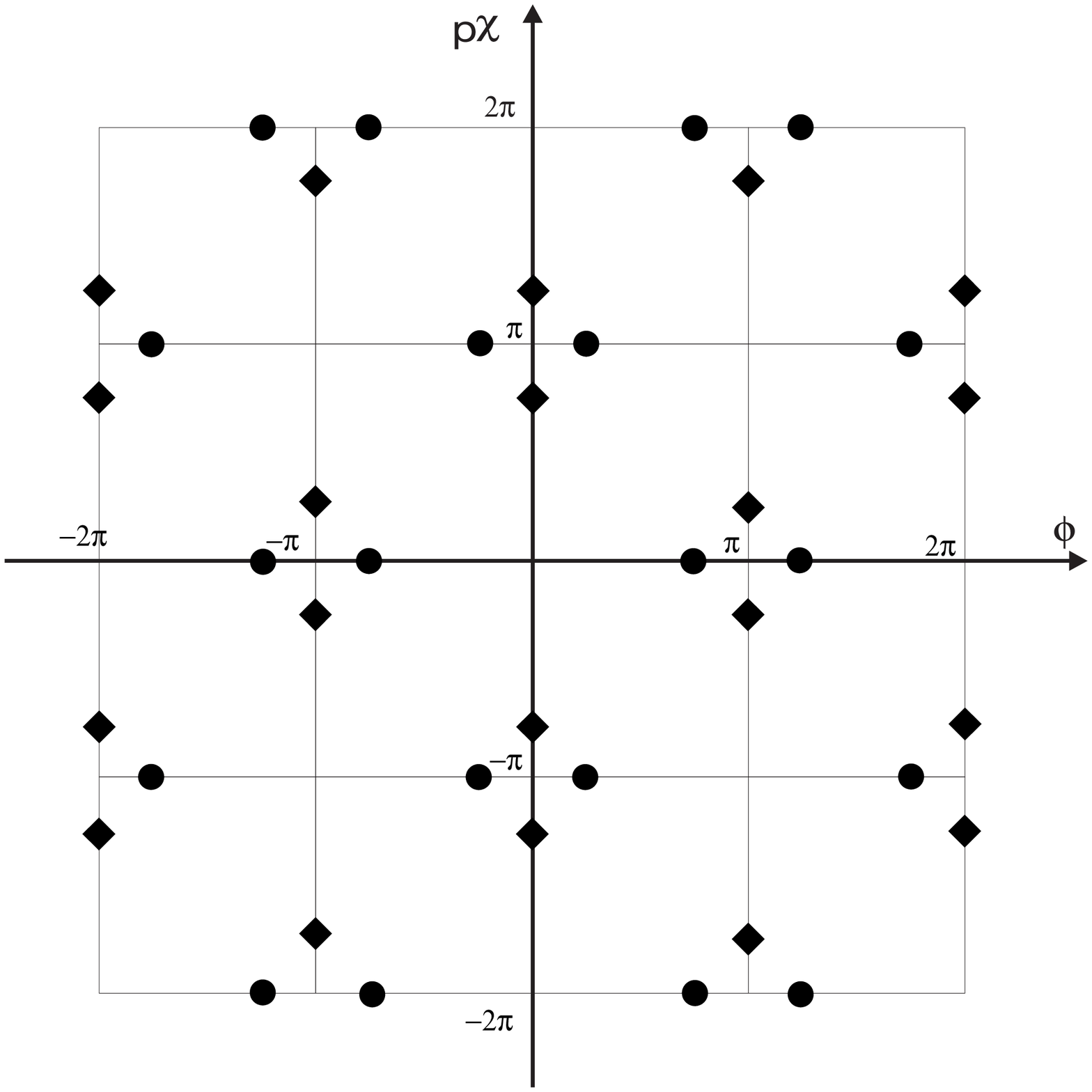}
\vspace{0.3cm}
\caption{The minima of model 2 potential. The  minima ${v}$ are represented
by balls and ${w}$ by losangles.}
\end{figure}

To obtain the most general solution to the first-order equations (\ref{dm21}) and
(\ref{dm22}), we integrate the ordinary differential equation
\be 
\frac{d\phi}{d\chi}=-r\frac{1}{p}\csc(\phi)\csc(p\,\chi)-
\frac{1}{p}\cot(\phi)\cot(p\,\chi) \nonumber
\ee
which admits the integrating factor $\sin^{-\sigma}(p\,\chi)$, with
$\sigma=1+1/p^2$ that determine all the orbits as the family of curves 
\be
\label{orb2}
\cos(\phi)=F(\chi)=\frac{1}{p}\sin^{\sigma-1}(p\,\chi)\left[C+r\,
\int\sin^{-\sigma}(p\,\chi)\,d\chi\right]
\ee
where $C$ is a real integration constant. In Fig.~10, we depict behavior
of a particular orbit with $C$, in the family (\ref{orb2}), where the critical
values are $C_{1}\approx-2.734$ and $C_{2}\approx2.734$, for $p=1/2$ and $r=1/2$.
\begin{figure}[ht]
\includegraphics[{height=6cm}]{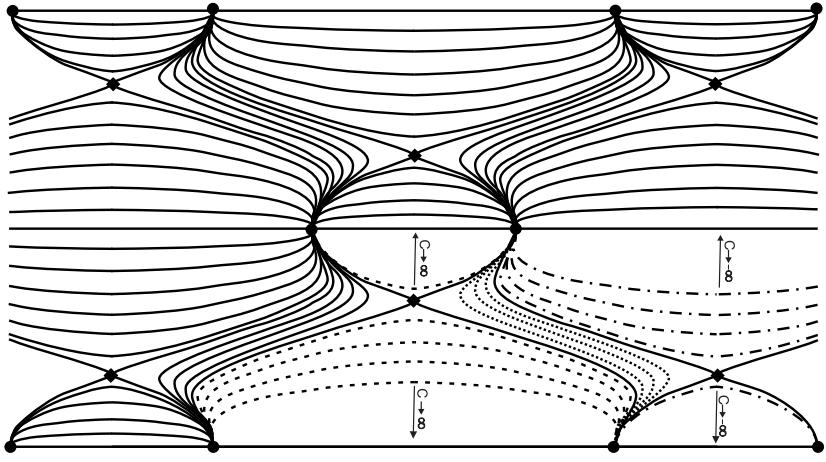}
\vspace{0.3cm}
\caption{Orbits given for: $-\infty<C<C_{1}$ (dash-dotted line),
$C=C_{1}$ (solid line between dash-dotted and dotted lines), $C_{1}<C<C_{2}$
 (dotted line), $C=C_{2}$ (solid line between dotted and dashed lines)
$C_{2}<C<\infty$ (dashed line).}
\end{figure}

There are five different BPS sectors connecting pairs of adjacent minima. They are
${v}{v}_{1}$  for $-\infty\leq C<C_{1}$ , ${v}{w}_{1}$ for $C=C_{1}$,
${v}{v}_{3}$  for $C_{1}<C<C_{2}$, ${v}{w}_{2}$ for $C=C_{2}$, and
${v}{v}_{2}$ for $C_{2}<C\leq\infty$. The tensions of the BPS states are 
$t_{vv_{1}}=2\,|\eta\,[\alpha\,r-\sin(\alpha)]|$,
$t_{vv_{2}}= 2\,|\eta\,[(\pi-\alpha)\,r+\sin(\alpha)]|$,
$t_{vv_{3}}=\frac12(t_{vv_{1}}+t_{vv_{2}})$, $t_{vw_{1}}=\frac12\, t_{vv_{1}}$,
and $t_{vw_{2}}=\frac12\, t_{vv_{2}}$.

Using the curve (\ref{orb2}), the BPS equation (\ref{dm22}) decouples,
and can be rewritten in the form, for $C\ne\pm\infty,$
\be
\label{dm22a}
\frac{d\chi}{dx}=-\eta\,p\,\sin(p\,\chi)\bigl[1-F^2(\chi)\bigr]^{\frac12}
\ee

We have been unable to obtain the general solutions of the above equation
(\ref{dm22a}). For this reason, we present some numerical solutions for the several
BPS sectors.

Firstly, we investigate the existence of kinks in the ${v}{v}_{1}$ and ${v}{v}_{2}$
sectors. Two horizontal adjacent minima are connected by the orbit $p\,\chi=k\pi$,
for $C=-\infty$ and $C=\infty$. In this case, the BPS equations (\ref{dif1})
and (\ref{dif2}) reduce to
\ben
\label{dif12}
\frac{d\phi}{dx}&=&\eta\,\bigl[r\pm\cos(\phi)\bigr]
\\
\label{dif22}
\frac{d\chi}{dx}&=&0
\een
The potential (\ref{v3}) becomes
\be
\label{vpm}
V_{\pm}=\frac12\eta^2\,\bigl[r\pm\,\cos(\phi)\bigr]^2
\ee
These potentials can be obtained from Eq.~(\ref{dsgg}) with
$\lambda=\frac{\eta}{2}\sqrt{1-r^2}$, $v=1/2$, and
$\alpha=\sqrt{\frac{1-r}{1+r}}$ or $\sqrt{\frac{1+r}{1-r}}$, for $V_{+}$ or
$V_{-}$, respectively. The solutions are given from Eq.~(\ref{phi1}) and
Eq.~(\ref{phi2}). For the potential $V_{+}$ the kink and antikink
solutions are:
\be
\chi_{vv_{1}}(x)=\chi_{vv_{2}}(x)=2n\frac{\pi}{p}
\ee
and
\be
\label{fi3}
\phi_{vv_{1}}(x)=(2n+1)\pi\pm
{2}\arctan\bigl[\sqrt{\frac{1-r}{1+r}}\tanh(\frac{\eta}{2}
\sqrt{1-r^2}\,x)\bigr]
\ee
and 
\be
\label{fi4}
\phi_{vv_{2}}(x)=2n\pi\pm
2\arctan\bigl[\sqrt{\frac{1+r}{1-r}}\tanh(\frac{\eta}{2}
\sqrt{1-r^2}\,x)\bigr]
\ee
For the potential $V_{-}$ the procedure is similar. We get
\be
\chi_{vv_{1}}(x)=\chi_{vv_{2}}(x)=(2n+1)\frac{\pi}{p}
\ee
and
\be
\label{fi5}
\phi_{vv_{1}}(x)=2n\pi\pm
{2}\arctan\bigl[\sqrt{\frac{1-r}{1+r}}
\tanh(\frac{\eta}{2}\sqrt{1-r^2}\,x)\bigr]
\ee
and 
\be
\label{fi6}
\phi_{vv_{2}}(x)=(2n+1)\pi\pm
{2}\arctan\bigl[\sqrt{\frac{1+r}{1-r}}
\tanh(\frac{\eta}{2}\sqrt{1-r^2}\,x)\bigr]
\ee

The numerical solutions are presented in figs.~11 and 12,
for $-\infty<C<C_{1}$ and $C_{2}<C<\infty$, respectively.

\begin{figure}[ht]
\includegraphics[{height=6cm}]{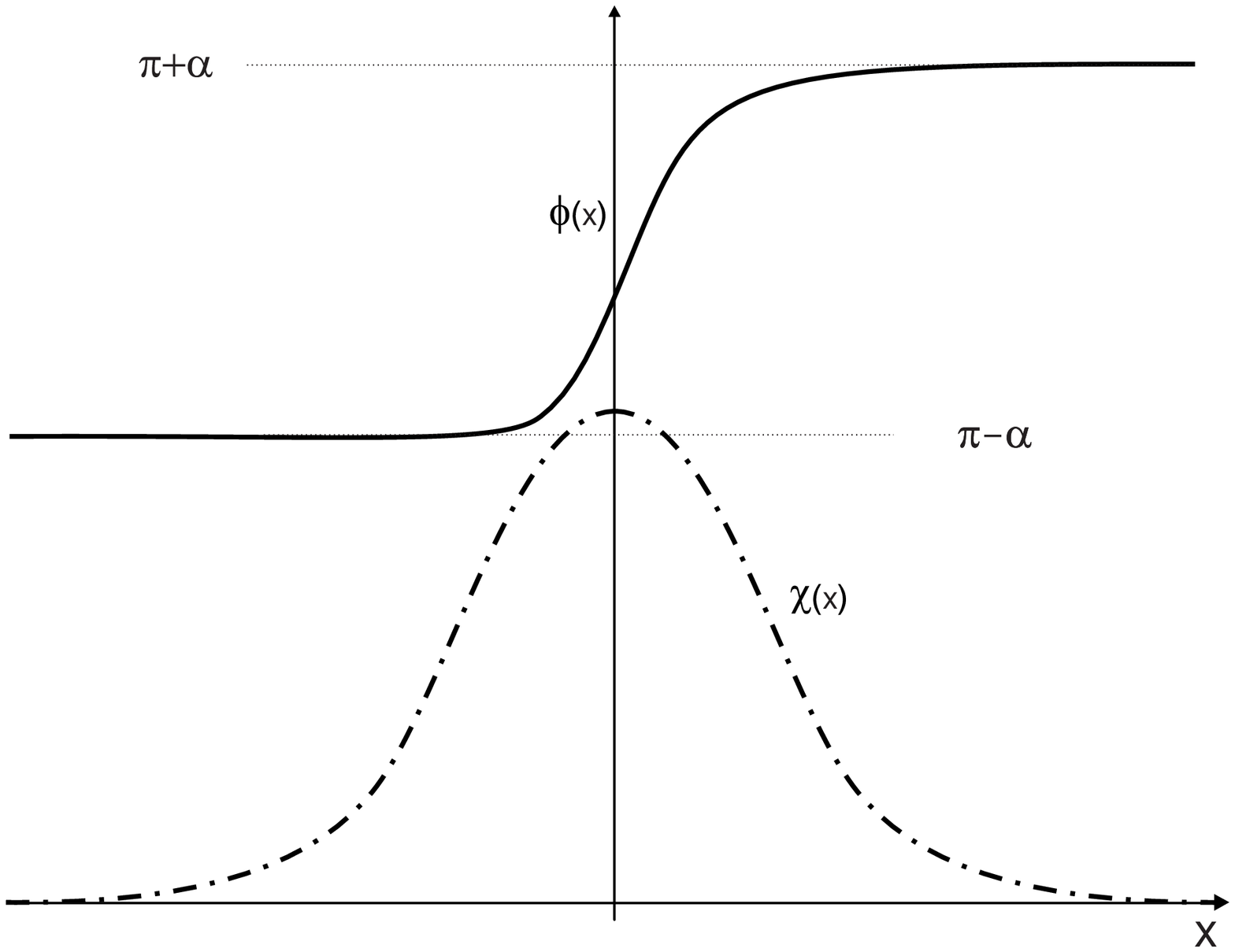}
\vspace{0.3cm}
\caption{Numerical solutions for ${\sl v}{\sl v}_{1}$ sector.}
\end{figure}

\begin{figure}[ht]
\includegraphics[{height=6cm}]{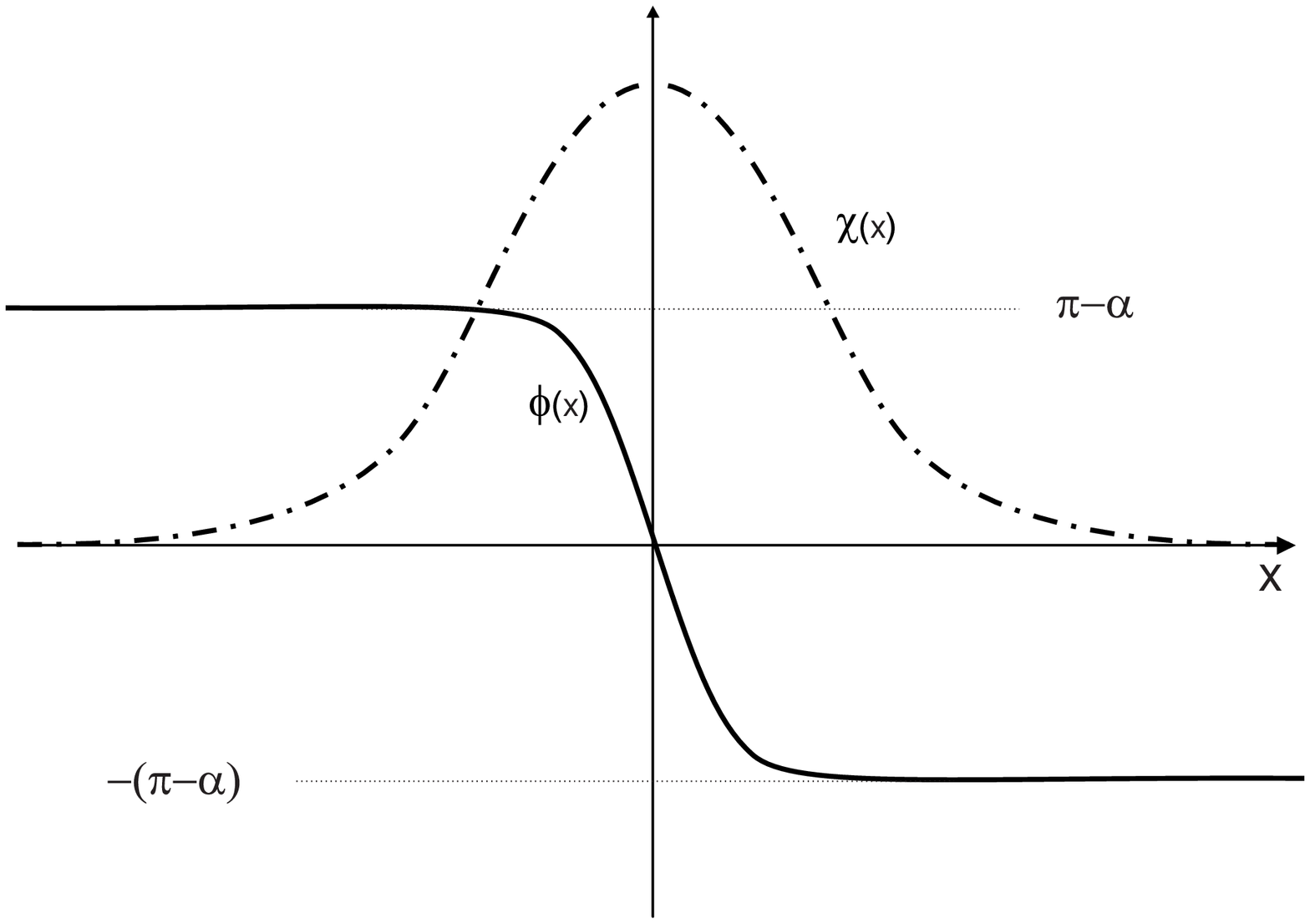}
\vspace{0.3cm}
\caption{Numerical solutions for ${\sl v}{\sl v}_{2}$ sector.}
\end{figure}

Kinks in the ${v}{w}_{1}$, ${v}{v}_{3}$ and ${v}{w}_{2}$ sectors are displayed
numerically in figs.~13, 14, and 15, for $C=C_{1}$, $C_{1}<C<C_{2}$, and $C=C_{2}$,
respectively.

\begin{figure}[ht]
\includegraphics[{height=6cm}]{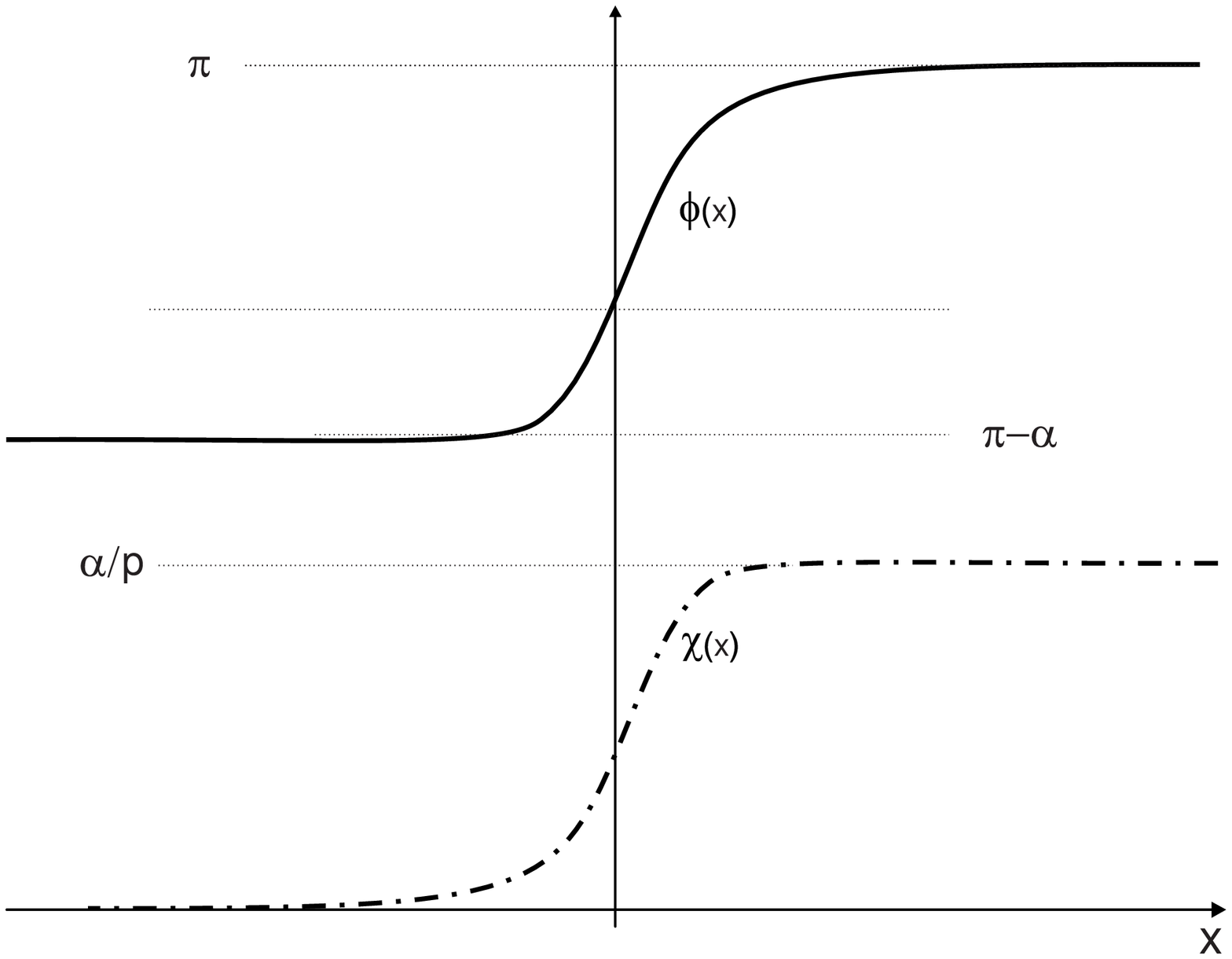}
\vspace{0.3cm}
\caption{Numerical solutions for ${\sl v}{\sl w}_{1}$ sector.}
\end{figure}

\begin{figure}[ht]
\includegraphics[{height=6cm}]{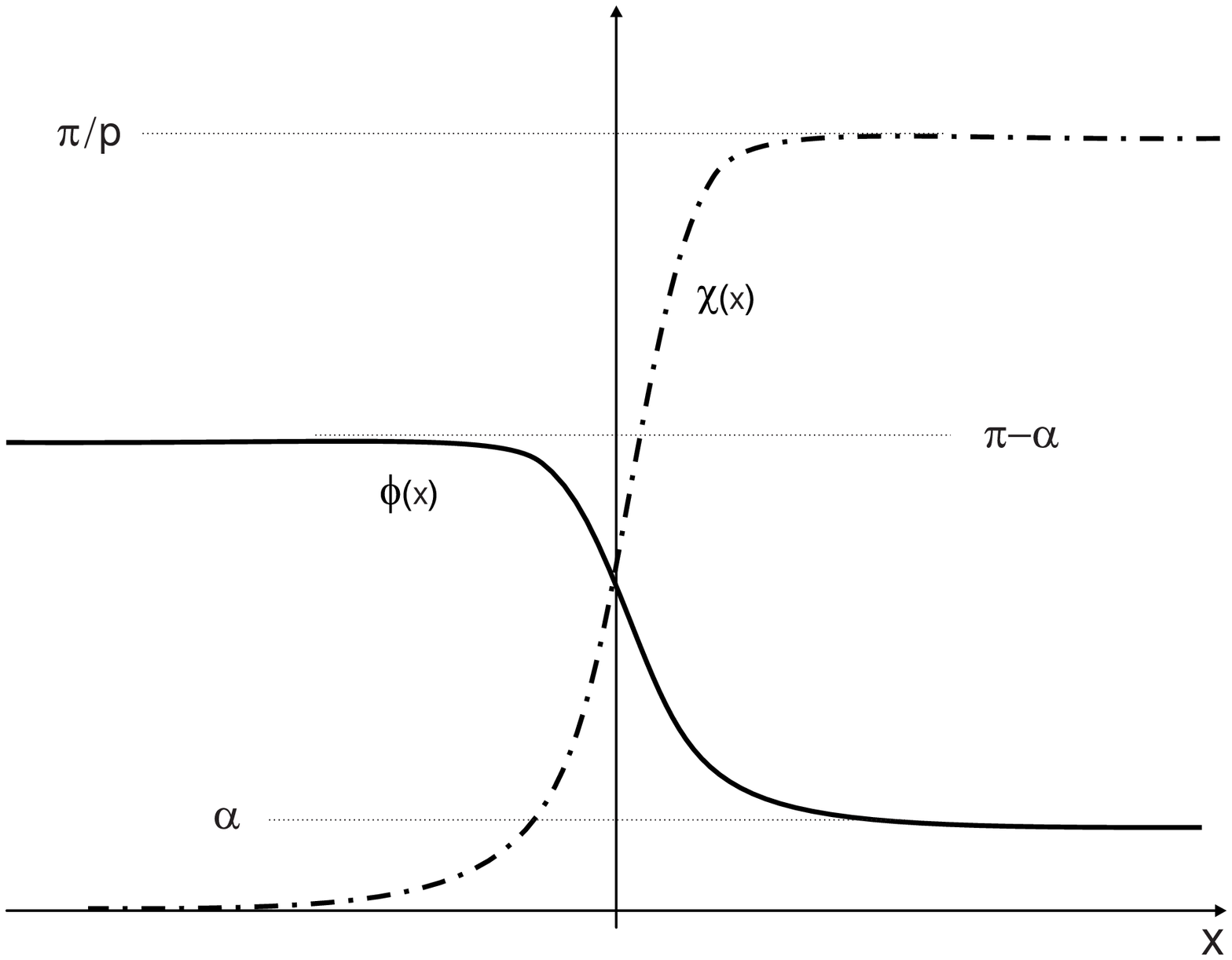}
\vspace{0.3cm}
\caption{Numerical solutions for ${\sl v}{\sl v}_{3}$ sector.}
\end{figure}

\begin{figure}[ht]
\includegraphics[{height=6cm}]{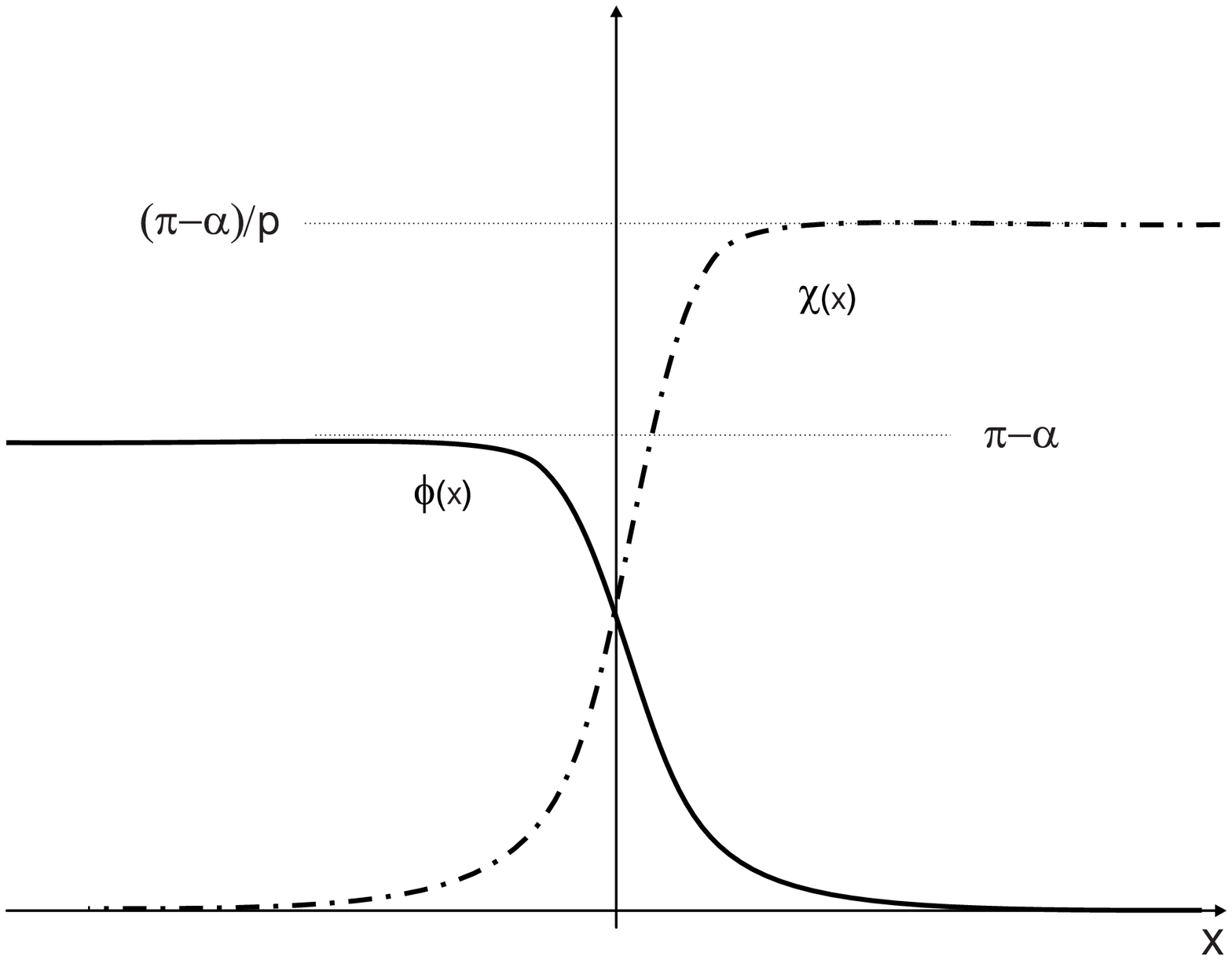}
\vspace{0.3cm}
\caption{Numerical solutions for ${v}{w}_{2}$ sector.}
\end{figure}

\subsection{Model 3}

The third model is given by the potential (\ref{v3}) with $r\geq 1$.
There is a set of minima at the points 
$u_{n,m}=(2n\,\pi\, ,\,(2m+1)\,\pi/p)$ and $v_{n,m}=((2n+1)\,\pi\, ,\,2m\,\pi/p)$.
In Fig.~16 we show these minima in the plane $(\phi,p\,\chi)$.

\begin{figure}[ht]
\includegraphics[{height=6cm}]{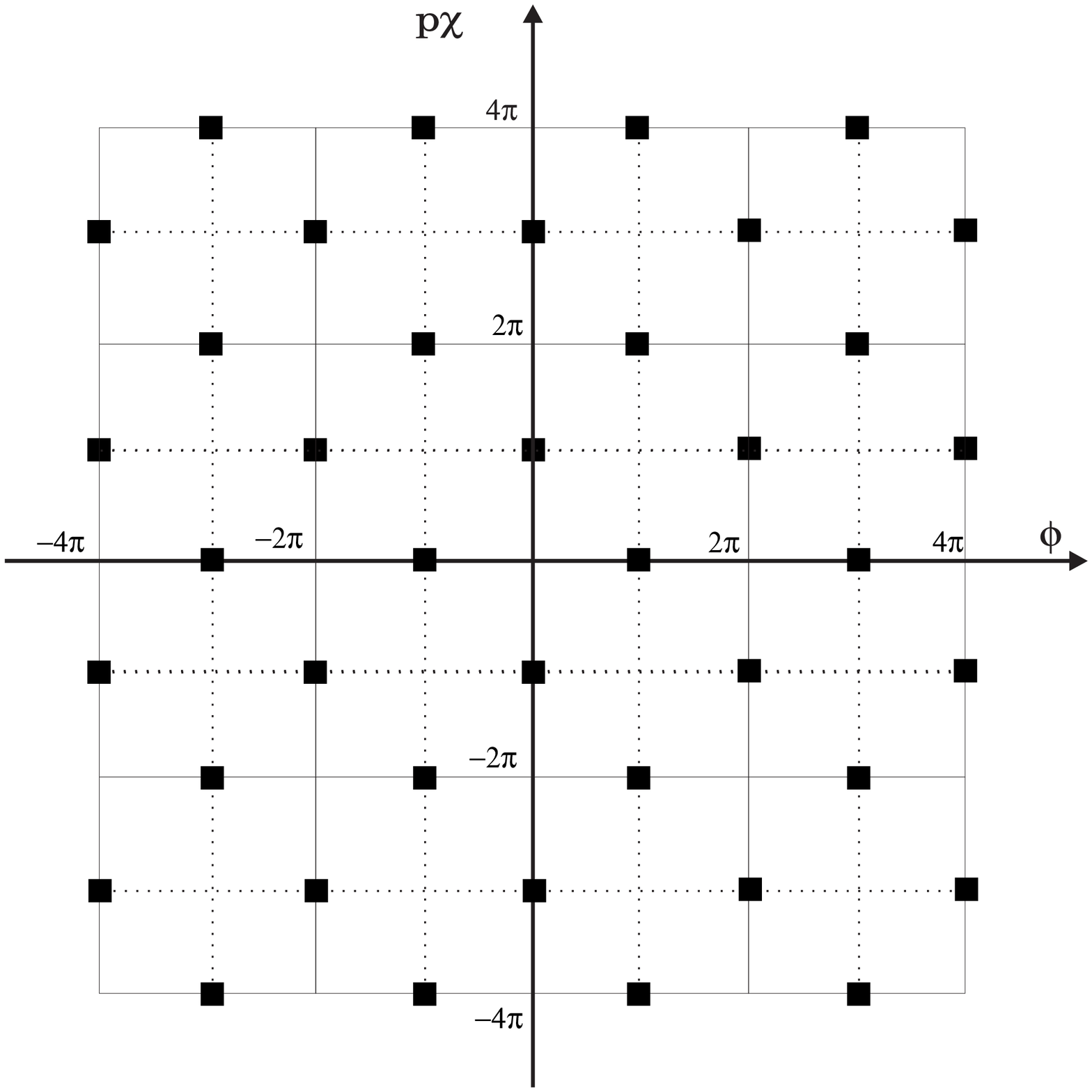}
\vspace{0.3cm}
\caption{The set of minima of the model 3.}
\end{figure}

Here, the first-order equations are (\ref{dm21}) and (\ref{dm22}). Thus, the orbits
are described by (\ref{orb2}).
There are two kinds of BPS sectors connecting pairs of adjacent minima in this case,
namely $uu$ or $vv$ for adjacent horizontal minima, and $uv$ for adjacent diagonal minima.
The tensions of the BPS states are $t_{uu}=t_{vv}=|\eta|\,2\,\pi\,r$ and
$t_{uv}=\frac12\,t_{vv}$.

In general, we could not obtain analytical solutions of Eq.~(\ref{dm22a}).
For this reason, we present some numerical solutions in the two BPS sectors.

Kinks in the $uu$ or $vv$ sectors are found for $C\to\infty$, connecting two horizontal
adjacent minima, with the orbit as a straight line $p\chi=k\pi$. This reduces the
potential (\ref{v3}) to the form (\ref{vpm}). In this case, the BPS equations
(\ref{dm21}) and (\ref{dm22}) have the solutions
\be
\chi(x)=k\frac{\pi}{p}
\ee
and  
\be
\label{fi31}
\phi(x)=k\pi\pm
{2}\arctan\Biggr[\sqrt{\frac{r+(-1)^k}{r-(-1)^k}}
\tanh\Bigr(\frac{\eta}{2}\sqrt{r^2-1}\,x\Bigl)\!\!\Biggl].
\ee
Particularly, for $r=1$ the solutions take the forms:
\be
\chi(x)=2n\frac{\pi}{p}, \,\,\,\,\,\, \phi(x)=
\pm{2}\arctan\bigr(\eta x\bigl)+k\pi
\ee
or
\be
\chi(x)=(2n+1)\frac{\pi}{p}, \,\,\,\,\,\,
\phi(x)=\pm{2}\,{\rm {arccot}}\bigr(\eta x\bigl)+k\pi
\ee

Some numerical solutions in the ${v}{u}$ sector, for $C=10$ and $p=\sqrt{3}/3$,
are depicted in Fig.~17.

\begin{figure}
\includegraphics[{height=6cm}]{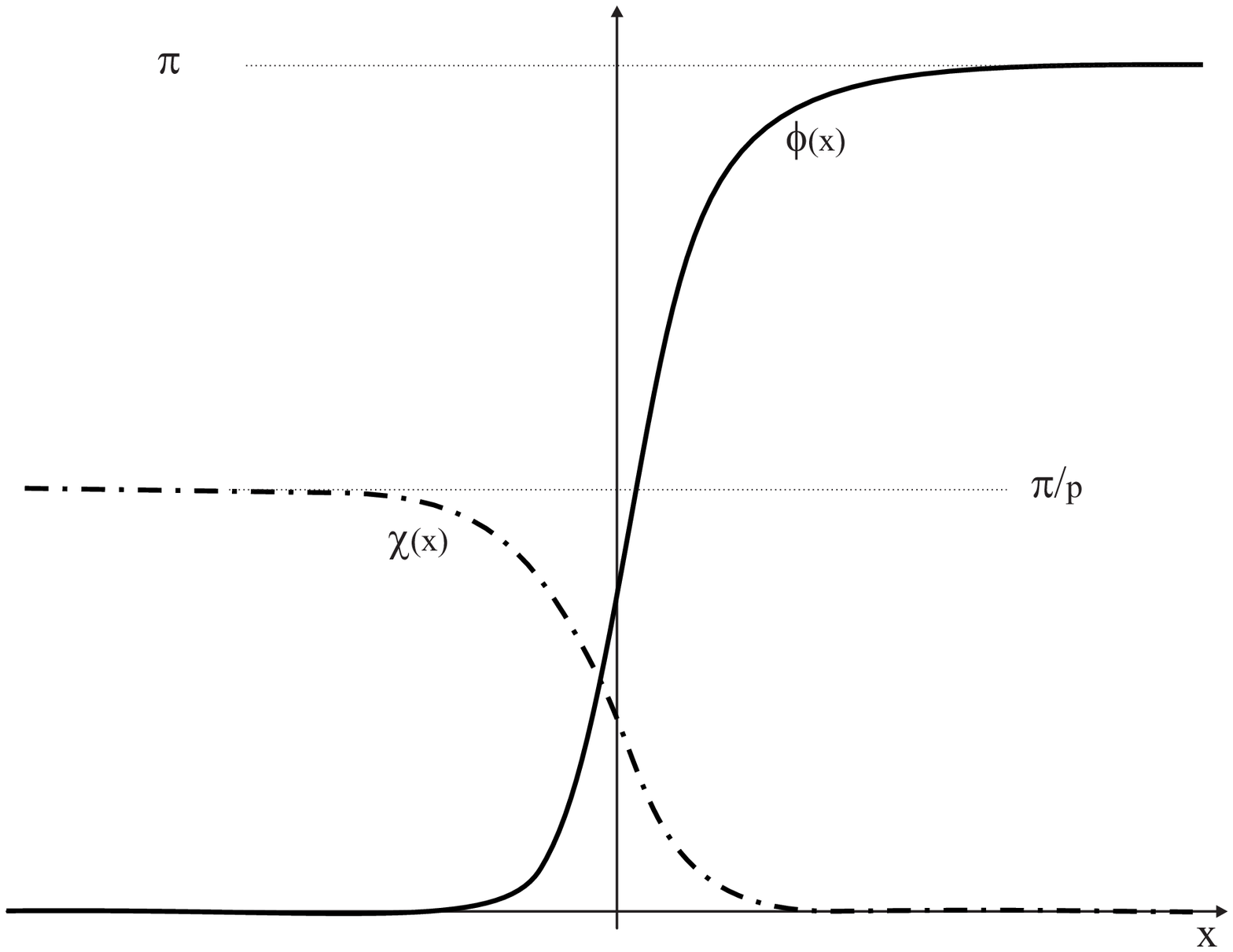}
\vspace{0.3cm}
\caption{Numerical solutions for ${u}{u}_{2}$ sector.}
\end{figure}

\section{Non-BPS solutions}
\label{nbps}

In this section we search for non-BPS solutions in the models introduced
in the former Sec.~{\ref{cla}}. To identify the non-BPS sectors, 
we investigate the equations of motion corresponding to each one
of the three models separately.

\subsection{Model 1}
\label{m11}

From the potential (\ref{v3}), with $r=0$, we obtain the equations of motion
\be
\label{dif50}
\frac{d^{2}\phi}{dx^{2}}=\frac12\eta^{2}\,\sin(2\,\phi)\bigl[p^2-(1+p^2)\,
\cos^{2}(p\,\chi)\bigr]
\ee
and
\be
\label{dif60}
\frac{d^{2}\chi}{dx^{2}}=\frac12\eta^{2}\,\sin(2p\,\chi)\bigl[p^2-(1+p^2)\,
\cos^{2}(\phi)\bigr]
\ee

There are straight-line orbits connecting two adjacent vertical and horizontal
${w}$ minima. The vertical orbit reduces the equations of motion (\ref{dif50})
and (\ref{dif60}) to
\be
\label{dif51}
\frac{d^{2}\phi}{dx^{2}}=0
\ee
and
\be
\label{dif61}
\frac{d^{2}\chi}{dx^{2}}=\frac12\eta^{2}\,p^3\,\sin^{2}(2p\,\chi)
\ee
The potential (\ref{v3}) reduces to (\ref{vc1}), with the kink and antikink solutions
\ben
\phi_{vv_{1}}(x)&=&k\pi \nonumber\\
\chi_{vv_{1}}(x)&=&\pm\frac{1}{p} \arccos\bigl[\pm\tanh(\eta\,p^2 x)\bigr])+k\pi \nonumber
\een
The horizontal orbit reduces the equations of motion (\ref{dif50}) and 
(\ref{dif60}) to
\ben
\label{df12}
\frac{d^{2}\phi}{dx^{2}}&=&\frac12\eta^2\,p^2\sin(2\,\phi)
\\
\label{df22}
\frac{d^2\chi}{dx^2}&=&0
\een
The potential (\ref{v3}) becomes $V=\frac12\eta^2\,p^2\,\sin^{2}(\phi)$,
with the kink and antikink solutions
\ben
\chi_{vv_{2}}(x)&=&(2k+1)\,\frac{\pi}{2p} \\
\phi_{vv_{2}}(x)&=&\pm \arccos\bigl[\pm\tanh(\eta\,q x)\bigr])+k\pi
\een

It is interesting to notice that if one considers the plane rotator model to describe
the open states in the molecule of deoxyribonucleic acid (DNA), in Ref.~\cite{Zh89}
the author has obtained the coupled sine-Gordon equations
\ben
(1-D^2/C_{0}^{2})\frac{d^2{\sl u}}{d\xi^2}&=&\frac{1}{{l}^2}\sin({u})+
\frac{2}{d^2}\sin({\frac12 u})\cos({\frac12 v}) \nonumber\\
(1-D^2/C_{0}^{2})\frac{d^2{v}}{d\xi^2}&=&\frac{Q}{{l}^2}\sin({v})+
\frac{2}{d^2}\sin({\frac12 v})\cos({\frac12 u}) \nonumber
\een
with ${u}=\varphi+\varphi'$ and ${v}=\varphi-\varphi'$, where D, $C_{0}$, ${l}$,
$Q$ and $d$ are physical constants, and $\varphi$ and $\varphi'$ describe the angular
displacements of two plane base-rotators. On the other hand, the equations of motions
(\ref{dif50}) and (\ref{dif60}) can be rewritten in the form
\ben
\label{dif52}
\frac{d^{2}\phi}{dx^{2}}&=&\frac12\eta^{2}\,\Bigl[\frac12 (1-p^2)\sin(2\,\phi)-
\frac12 (1+p^2)\sin(2\,\phi)\,\cos(2p\,\chi)\Bigr]
\\
\label{dif62}
\frac{d^{2}\chi}{dx^{2}}&=&\frac12\eta^{2}\,\Bigl[\frac12 (1-p^2)\sin(2p\,\chi)-
\frac12 (1+p^2)\sin(2p\,\chi)\,\cos(2\,\phi)\Bigr]
\een
We notice that this model can be used as an alternative to the DNA model;
this will be further examined in another work.

\subsection{Model 2}
\label{m22}

From the potential (\ref{v3}), with $0<r<1$, we obtain the equations of motion
\ben
\label{dif7}
\frac{d^{2}\phi}{dx^{2}}&=&\frac12\eta^{2}\,\sin(\phi)\bigl[p^2\cos(\phi)
\sin^{2}(p\,\chi)
-p^2\,\cos(q\,\chi)[r+\cos(p\,\phi)\cos(q\,\chi)]\bigr]
\\
\label{dif8}
\frac{d^{2}\chi}{dx^{2}}&=&\frac12\eta^{2}\,p\,\sin(p\,\chi)
\bigl[p^2\cos(p\,\chi)\sin^{2}(\phi)
-\cos(\phi)[r+\cos(\phi)\cos(p\,\chi)]\bigr]
\een

There are straight-line orbits, $\phi=k\,\pi$, connecting two adjacent vertical
${w}$ minima. The orbit reduces the potential (\ref{v3}) to
$V_{\pm}=\frac12\eta^2\,\bigl[r\pm\,\cos(p\,\chi)\bigr]^2$.
For the potential $V_{+}$, the solutions are
\be
\phi_{ww_{(1,2)}}(x)=k\pi, \,\,\,\,\,\,\, \,\,\,\chi_{ww_{(1,2)}}(x)=
\frac{1}{p}\,\phi_{vv_{(1,2)}}(x)
\ee
where $\phi_{vv_{(1,2)}}$ are given by (\ref{fi3}) and (\ref{fi4}), respectively.
For the potential $V_{-}$, the  solutions are
\be
\phi_{ww_{(1,2)}}(x)=(2k+1)\pi, \,\,\,\,\,\,
\chi_{ww_{(1,2)}}(x)=\frac{1}{p}\,\phi_{vv_{(1,2)}}(x)
\ee
where $\phi_{vv_{(1,2)}}$ are given by (\ref{fi5}) and (\ref{fi6}), respectively.

\subsection{Model 3}
\label{m33}

From the potential (\ref{v3}), with $r\geq1$, we obtain the equations of motion
(\ref{dif7}) and (\ref{dif8}). There are straight-line orbits, $\phi=k\,\pi$,
connecting two adjacent vertical $u$ or $v$ minima. The orbit reduces the
potential (\ref{v3}) to
\be
V=\frac12\eta^2\,\bigl[r+(-1)^k\,\cos(p\,\chi)\bigr]^2.
\ee
The solutions are
\be
\phi(x)=k\pi,\,\, \chi(x)=\frac{1}{p}\,{\wt\phi}(x),
\ee
where ${\wt\phi}$ is given by eq.~(\ref{fi31}). 

The non-BPS solutions may be unstable, and may decay into stable BPS
solutions. The decay of non-BPS states is out of the scope of this work, but it
can be done following for instance the lines of Ref.~{\cite{riazi}}.

\section{Comments and conclusions}
\label{con}

In the present work we have investigated several models described by one and
by two real scalar fields. The main investigations concern the search for
BPS states, that is, for topological solutions that solve first order
differential equations. These solutions minimize the energy to the
Bogomol'nyi bound, which is given solely in terms of the superpotential,
and the asymptotic value of the corresponding field configurations.

The search for topological solutions is done at the classical level, and
we have payed special attention to models
introduced in Sec.~{\ref{gen}}, some of them further explored in
Sec.~{\ref{cla}}. These investigations have shown how to write
the double sine-Gordon model in terms of a superpotential, and to deal with
the large and small kinks as BPS states, as solutions to first order
differential equations. We have also investigated another model,
decribed by two real scalar fields, defined by the potential of Eq.~(\ref{v3}).
It is similar to the model first investigated in Ref.~{\cite{95}}, although here
we deal only with periodic interactions.

The classical investigations that we have developed are of interest in applications to
nonlinear science, as for instance in the line of the work presented in
Refs.~{\cite{96,99,00,01}}, where one uses field theory models to mimic
nonlinear interactions in polymeric chains. For instance, the model 1 may
be used as an alternative to the $\phi^4$ model that is standardly considered
to mimic the polyacethylene (PA) chain, where Peierls instability appears due
to single and double bound alternation in carbon atoms along the
chain. The standard scenario leads to the very nice picture in which the
distance carbon-carbon is mapped to the $\phi^4$ model with spontaneous
symmetry breaking. The need for spontaneous symmetry breaking is to reproduce
the two degenerate states, which describe single-double and double-single bond
alternations in the trans or zig-zag PA chain. Of course, this picture can become
more interesting if one adds fermions to the system, via the standard Yukawa
coupling. The alternative that we propose is to mimic the PA chain with
model 1, which is very much similar to the $\phi^4$ model at the classical
level. We hope to explore this and similar ideas in the near future.

A direct motivation that follows from the present investigations concerns
the inclusion of fermions, to see how the fermions change the scenario
we have just obtained. Another motivation concerns generalization of the
field theory models to the case of complex fields, which may
give rise to models engendering the continuum $U(1)$ symmetry. In this
new scenario the models admit the presence of vortices, global and local,
depending on the gauging of the global symmetry that appears when one
changes the real field to a complex one. Furthermore, the two-field model
investigated in Sec.~{\ref{cla}} is of direct use to describe interactions
between two BEC's, and the corresponding vector solution describes the
interface between the interacting condensates, as a good alternative to
the recently proposed model of Ref.~{\cite{cha}}, because in our model the
center of the interface is asymmetric, containing different quantities
of each one of the two condensates that form the interface. This is more
realist than the case studied in \cite{cha}, where the center of the interface
contains equal portions of each one of the two condensates. All the models
here introduced are also of interest to pattern formation in spatially
extended, periodically forced systems, governed by the Ginzburg-Landau
equation. This is a new route, different from the one proposed in \cite{96},
and further investigated in \cite{99,00,01}, which deals with solitons
in ferroelectric cristals, in polyethylene, and in Langmuir films. The
Ginzburg-Landau equation is appropriate for investigating fronts and
interfaces, and their contributions to pattern formation. These and other
specific issues are presently under investigation, and we hope to report
on them in the near future.

\acknowledgments

The authors would like to thank PROCAD/CAPES and PRONEX/FAPESQ/CNPq for financial
support. DB thanks CNPq for partial support, and RM thanks CAPES for a fellowship.

\end{document}